\documentclass[aip,graphicx,jcp]{revtex4-1}
\usepackage[utf8]{inputenc}
\usepackage[T1]{fontenc}
\usepackage{amsmath,amssymb,bm}
\usepackage{graphicx}
\usepackage{booktabs}
\usepackage{siunitx}
\usepackage{subcaption}
\usepackage[margin=2.5cm]{geometry}
\usepackage{float}
\usepackage{multirow}
\usepackage{xcolor}
\usepackage[version=4]{mhchem}
\usepackage[colorlinks=true,linkcolor=blue,citecolor=blue,urlcolor=blue]{hyperref}
\usepackage{tikz}

\newcommand{\eps}{\varepsilon}
\newcommand{\epsinf}{\varepsilon^{\infty}}
\newcommand{\epszero}{\varepsilon_{0}}
\newcommand{\epsstilde}{\tilde{\varepsilon}_{\mathrm{s}}}
\newcommand{\epss}{\varepsilon_{\mathrm{s}}}
\newcommand{\Deps}{\Delta\varepsilon}
\newcommand{\epspp}{\varepsilon''}
\newcommand{\Mtot}{\bm{M}_{\mathrm{tot}}}
\newcommand{\Mdef}{\bm{M}_{\mathrm{def}}}
\newcommand{\Mmat}{\bm{M}_{\mathrm{mat}}}
\newcommand{\tauavg}{\langle\tau\rangle}

\begin{document}

\title{Multiscale Computational Study of the Dielectric Response
  of Semi-Crystalline Polyethylene with Chemical Defects} %

\author{Roshal Perepadan Shaju}
\author{Guido Roma}
\email{guido.roma@cea.fr}
\affiliation{Universit\'e Paris-Saclay, CEA, Service de Recherches en Corrosion et Comportement des Mat\'eriaux, SRMP, 91191 Gif sur Yvette, France}
\author{Xavier Colin}
\affiliation{PIMM, Arts et M\'etiers Institute of Technology, CNRS, CNAM, HESAM University, 151 Boulevard de L’H\^opital, 75013 Paris}

\date{\today}

\begin{abstract}
Polymers are widely used as functional insulating materials, but predicting their dielectric behavior is challenging because multiple mechanisms, acting across different spatial and temporal scales, contribute to their response. In polyethylene (PE), one of the most common polymers, electronic, atomic, and mesoscale dynamics must all be considered.

In this work, semicrystalline models of polyethylene were developed and investigated using a multiscale approach. Quantum simulations were employed to describe electronic and vibrational contributions to the dielectric response, while classical molecular dynamics simulations captured the behavior of polymer chains at room temperature and frequencies down to the 10--100~MHz range.

The study focuses on the impact of radio-oxidation defects on the dielectric properties of polyethylene. Quantum calculations reveal how electronic and vibrational contributions depend on the local atomic environment surrounding these defects. Comparison with molecular dynamics results highlights the influence of temperature on the static dielectric constant. Structural analyses further assess the effect of each defect type on PE crystallinity.

Analysis of dipolar correlation functions shows that different defects interact with one another and with the polymer matrix in distinct ways, affecting permittivity and dielectric loss peaks. For all defect types, defect–defect interactions provide the largest contribution to dielectric response. While most oxidized groups exhibit positive coupling with the PE matrix, alcohol defects display a negative cross-correlation, partially offsetting their impact on the dielectric properties.

Our results also show that ketone groups produce the largest dielectric loss between the defects studied.

\end{abstract}


\maketitle %

\section{Introduction}
\label{sec:intro}

\begin{figure}[htbp]
  \centering
  \includegraphics[width=0.85\textwidth]{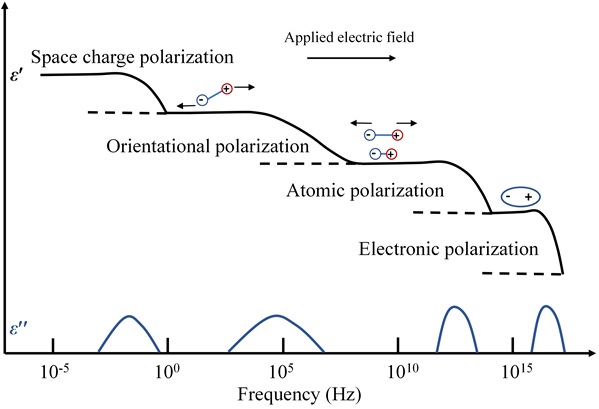}
  \caption{Schematic view of the polarisability of a polymer as a function of log frequency (Hz) showing the relaxation and resonance regions for dipolar, atomic, and electronic polarisation. Adapted from Kremer and Sch\"onhals\cite{Schoenhals2003}.}
  \label{fig:polarization_mechanism}
\end{figure}

	Polyethylene (PE) is a widely used polymer dielectric for electrical insulation in high-voltage cable systems and harsh-service environments (e.g., nuclear power plants) due to its low relative permittivity and exceptionally low dielectric loss \cite{bur1985dielectric,Conklin1964JAP}.  The dielectric properties of a polymer are frequency-dependent and originate from several polarization mechanisms across different frequency regimes. Fundamentally, the relative permittivity under an external electric field is determined by several microscopic polarization mechanisms: (i)~electronic polarization (electron-cloud distortions), (ii)~ionic/vibrational polarization (displacement of atoms or segments), and (iii)~dipolar/orientational polarization (alignment of permanent dipoles) \cite{boyd1984strengths,huan2016advanced} (see Fig.~\ref{fig:polarization_mechanism}). 
In general, electronic polarization remains active at optical and higher frequencies, ionic polarization becomes active in the infrared–terahertz regime, and dipolar or orientational polarization contributes at lower frequencies, typically in the microwave-to-radio-frequency range (see Fig.~\ref{fig:polarization_mechanism}).
For non-polar polymers such as PE, the dielectric response is largely governed by the electronic polarization of the distorted valence electron clouds on the C--C and C--H bonds with a minor contribution from fast vibrational modes. Hence, the real part of the permittivity stays close to $\eps_r \approx 2.2$--$2.3$ over broad frequency ranges \cite{bur1985dielectric}, has minimal dielectric dispersion, and the loss factor (dielectric loss $\epspp(\omega)$) remains extremely low at normal temperatures.

Although ideal polyethylene chains carry no polar groups, chemical defects are formed as PE undergoes ageing. High-energy radiation (e.g.\ $\gamma$-rays, electron beams, X-rays) can trigger the breaking of C--H or C--C bonds producing radicals that react with \ce{O2} to form peroxy radicals and hydroperoxides, and subsequently oxygen-containing polar defects \cite{Ahn2021Polymers, DaCruz2016JAPS, Suraci2020ICD}. Such defects even at low concentrations can orient in an electric field, thus contributing a dipolar polarization component, which raises the static permittivity significantly. Experimental dielectric spectroscopy has shown that polar defects can produce additional relaxation processes and increase dielectric   loss \cite{gilchrist1987,stamboliev2010dielectric,ashcraft1976dielectric,sato1978,mujalrosas2015}. Theoretical and computational studies can enable an understanding of the molecular-level origins of dielectric properties \cite{umari2002ab,wang2014computational}. However, a key complication is that PE is semi-crystalline, consisting of crystalline lamell\ae\ separated by amorphous and interfacial regions. A chemical defect behaves differently depending on whether it is located in the crystalline region, the amorphous region, or at the interface between them. This affects the dielectric properties and makes the analysis more complex. Previous computational studies have indicated that chemical defects can alter the anisotropic dielectric response of crystalline PE and that morphology can strongly influence the effective permittivity in semi-crystalline polyolefins \cite{fukushima2019,misra2016,gedde1995polymer}. 

In this work, we study the frequency-dependent dielectric properties of PE using a multiscale approach that combines density functional perturbation theory (DFPT) and classical molecular dynamics (MD). First, DFPT is used to compute the high-frequency (electronic) dielectric tensor, $\epsinf$, and the static dielectric tensor including ionic (vibrational) contributions, $\epszero$, for a semi-crystalline lamellar slab model that explicitly contains crystalline, interfacial, and amorphous environments \cite{ferry2023}. We examine seven functional defects---namely alcohol, aldehyde, ketone, carboxylic acid, enone, ester, and vinylene---at two low concentrations (2 and 4~mol\,\%) distributed at different locations throughout the polymer structure. Second, long-time MD simulations capture the evolution of the morphology at finite-temperature and frequency-dependent permittivity via the dipole autocorrelation function, where the contributions from the orientational polarization dynamics are accounted for. From fluctuations in the dipole moment, we extract the static dielectric constant and the complex dielectric response $\eps(\omega) = \eps'(\omega) - i\epspp(\omega)$. The integration of these methods provides a comprehensive understanding of polymer dielectrics, enabling quantitative comparison with experiments, and supporting improved predictive models for polymers operating under high-field, high-stress conditions.

\section{Methodology}
\label{sec:methodology}

\subsection{DFPT Dielectric Response}
\label{sec:dfpt_method}

\subsubsection{Simulation Details}

A semi-crystalline PE is characterised by ordered crystalline lamell\ae\ separated by amorphous inter-lamellar layers \cite{gedde1995polymer}. We replicate the semi-crystalline structure of PE employing a lamellar slab model derived from previous studies by  Ahn~\emph{et al.} \cite{Ahn2022Macromolecules} and Ferry~\emph{et al.} \cite{ferry2023}. An orthorhombic lattice of crystalline lamell\ae\ composed of all-\emph{trans} chains features two free surfaces connected across a vacuum space by an alkyl chain that resembles an amorphous region. This configuration effectively captures the ordered crystalline, disordered amorphous, and interfacial regions that are characteristic of semi-crystalline polymers.

The simulation cell initially contains 50 \ce{CH2} units. The chemical defects are investigated at two low concentrations,  2~mol\,\% and 4~mol\,\%, which are modeled by replacing 1 and 2 \ce{CH2} units in the simulation cell, respectively. For a concentration of 4~mol\,\%, one defect was fixed at the crystalline--amorphous interface, while the second defect was placed in various locations in different configurations (spanning crystalline, interfacial, and amorphous regions) to evaluate the influence of the local environment and the potential defect--defect interaction.

For example, to model an alcohol defect (--OH), a hydrogen atom on a \ce{CH2} group was replaced with a hydroxyl group, resulting in a --CHOH-- motif in the polymer backbone. To model a ketone defect, a methylene group (\ce{-CH2-}) was replaced with a carbonyl group (\ce{C=O}) by removing two hydrogens and forming a double bond between carbon and oxygen. To emulate an aldehyde defect, the PE chain was broken and a --CHO group was introduced at one chain end, while the other was passivated as CH$_3$. All modified structures were fully relaxed using the same criteria as for the pristine cell prior to dielectric analysis. Thus, we quantify how each of the above defects impacts the dielectric permittivity of polyethylene with respect to its formation in different morphological regions.

Density Functional Theory (DFT) calculations were performed with \textsc{Quantum ESPRESSO} \cite{giannozzi2009}. Norm-conserving pseudopotentials and a plane-wave basis set with a kinetic energy cutoff of 80~Ry were used. The periodic simulation cell ($9.72 \times 6.95 \times 19.95$~\AA) contains 151 atoms (pristine PE), with a density of 0.88~g/cm$^3$ \cite{sharma2012}. A $\Gamma$-centered $2\times3\times2$ Monkhorst--Pack grid was used to sample the Brillouin zone. Geometry optimisations were performed until the residual forces were minimised to less than $10^{-4}$~Ry/Bohr. The optB86b + vdW  van der Waals density functional was used, as it is known to provide an improved description of dispersion interactions, which play a significant role in polymers such as PE because of their long-range intermolecular forces \cite{thonhauser2015spin,klimes2010chemical}. The calculated dielectric tensor components $\eps_{\alpha\beta}$ (in the frame of the crystalline axis) were recorded for the pristine system and for each configuration of defects.

\subsubsection{High-Frequency and Static Dielectric Constant}

From DFPT calculations, both the high-frequency dielectric tensor, $\epsinf$, and the static dielectric tensor, $\epszero$, can be calculated. We use $\epszero$ for the static dielectric constant calculated from DFPT at zero temperature to distinguish it from the static dielectric constant calculated from MD at finite temperature ($\epss$). We obtained $\epsinf$ and $\epszero$ for each defect-containing relaxed structure. DFPT calculations were performed at the $\Gamma$ point using the \texttt{ph.x} module of Quantum ESPRESSO with \texttt{epsil=.true.}, which enables the calculation of the macroscopic dielectric tensor for non-metallic systems. The corresponding $\Gamma$-point dynamical matrices were then analyzed using \texttt{dynmat.x}, which diagonalizes the dynamical matrix and provides access to the zone-centre vibrational modes relevant to the ionic contribution to the dielectric response. The high-frequency dielectric constant represents the purely electronic contribution to the dielectric response, i.e.\ the response of the electron cloud with the ionic positions held fixed. In contrast, the static dielectric constant corresponds to the zero-frequency limit of the dielectric response and includes, in addition to the electronic polarization, the lattice contribution arising from infrared-active zone-centre phonon modes. As a result, $\epszero$ accounts for both electronic and ionic polarization effects, whereas $\epsinf$ contains only the electronic part.

Accordingly, the static dielectric tensor can be written as

\begin{equation}
  \varepsilon^0_{\alpha\beta} = \varepsilon^{\infty}_{\alpha\beta} + \varepsilon^{\mathrm{ion}}_{\alpha\beta},
  \label{eq:eq1_tensor}
\end{equation}

where $\varepsilon^{\mathrm{ion}}_{\alpha\beta}$ denotes the ionic contribution associated with the zone-centre vibrational modes. In mode-resolved form, this contribution may be expressed as\cite{Rignanese_2001}

\begin{equation}
  \varepsilon^{0}_{\alpha\beta}
=
\varepsilon^{\infty}_{\alpha\beta}
+
\frac{4\pi}{\Omega_{0}}
\sum_{m}
\frac{S_{m,\alpha\beta}}{\omega_{m}^{2}},
  \label{eq:eq2_mode_resolved}
\end{equation}

where $\Omega_{0}$ is the unit-cell volume, $\omega_{m}$ is the frequency of the $m^{\text{th}}$ infrared-active phonon mode at the Brillouin-zone centre, and $S_{m,\alpha\beta}$ is the corresponding oscillator-strength tensor. The dielectric constants reported in this work are scalar averages of the diagonal components of the corresponding dielectric tensors unless otherwise noted, computed as
\[
\bar{\varepsilon} = \frac{\varepsilon_{xx} + \varepsilon_{yy} + \varepsilon_{zz}}{3}.
\]
Accordingly, both $\epsinf$ and $\epszero$ refer to the orientationally averaged values of the electronic and total static dielectric tensors, respectively.

\subsection{Molecular Dynamics Simulations}
\label{sec:md_method}

\subsubsection{Simulation Details}

Classical molecular dynamics (MD) simulations were carried out using LAMMPS \cite{LAMMPS}. The transferable OPLS-AA force field was employed to describe the PE chains and the polar defect groups \cite{jorgensen2023opls}. Initial configurations were generated using Moltemplate \cite{jewett2021} and consisted of a single linear PE chain containing 2000 methylene units. The PE chain was first constructed in an extended conformation and subsequently subjected to a melt--quench procedure to generate a semi-crystalline morphology. Energy minimization was performed using the conjugate-gradient algorithm. The minimized structure was then equilibrated in the melt state at 600~K using the NVT ensemble. Subsequently, the system was cooled to 300~K at a rate of 20~K$\cdot$ns$^{-1}$ under the NPT ensemble, using isotropic pressure coupling at 1 atm. 
The volume was equilibrated and stabilized at 300~K for 50~ns, followed by an additional 5~ns in the NVT ensemble. This equilibrated structure was then used as the initial configuration for production simulations, which were performed for 100--150~ns under NVT conditions.

Temperature and pressure were controlled using the Nos\'e--Hoover thermostat and barostat, with damping time constants of 2~ps and 10~ps, respectively. Long-range electrostatic interactions were treated using the particle--particle particle--mesh (PPPM) method with a real-space cutoff of 1.0~nm, while van der Waals interactions were described using a Lennard--Jones potential with a cutoff of 1.0~nm. Partial atomic charges for the polar defect groups were assigned explicitly using the OPLS-AA force-field parameters and are summarized in Table~\ref{tab:defect_charges}. In the present MD framework, the force field is non-polarizable, such that the atomic partial charges remain fixed throughout the simulation and the dielectric response arises solely from the configurational fluctuations of these permanent charge distributions. During the production runs, the total dipole moment was sampled every 100~fs. Accordingly, the total dipole moment of the simulation cell at time $t$ was computed as
\[
\bm{M}(t)=\sum_i q_i\,\bm{r}_i(t),
\]
where $q_i$ and $\bm{r}_i(t)$ are the partial charge and position vector of atom $i$, respectively. The time evolution of $\bm{M}(t)$ was then used to evaluate both the dipole-moment fluctuations and the dipole autocorrelation function employed in the dielectric analysis.

\begin{table}[htbp]
\centering
\caption{Partial atomic charges assigned to the PE backbone and selected polar defect moieties. Charges are given in units of the elementary charge, $e$. The subtotal corresponds to the contribution of each atomic type within the corresponding group.}
\label{tab:defect_charges}
\begin{tabular}{llccc}
\hline
Group/Defect & Atomic site & Multiplicity & Charge ($e$) & Subtotal ($e$) \\
\hline
Backbone & C(\ce{CH3})              & 1 & -0.180 & -0.180 \\
         & H(\ce{CH3})              & 3 &  \phantom{-}0.060 &  \phantom{-}0.180 \\
         & C(\ce{CH2})              & 1 & -0.120 & -0.120 \\
         & H(\ce{CH2})              & 2 &  \phantom{-}0.060 &  \phantom{-}0.120 \\
\hline
Alcohol  & C(\ce{CHOH})             & 1 &  \phantom{-}0.205 &  \phantom{-}0.205 \\
         & H(\ce{CHOH})             & 1 &  \phantom{-}0.060 &  \phantom{-}0.060 \\
         & O(\ce{OH})               & 1 & -0.683 & -0.683 \\
         & H(\ce{OH})               & 1 &  \phantom{-}0.418 &  \phantom{-}0.418 \\
\hline
Aldehyde & C(\ce{CHO})              & 1 &  \phantom{-}0.450 &  \phantom{-}0.450 \\
         & H(\ce{CHO})              & 1 &  \phantom{-}0.000 &  \phantom{-}0.000 \\
         & O(\ce{CHO})              & 1 & -0.450 & -0.450 \\
\hline
Ketone   & C(\ce{CH2-$\alpha$})     & 1 & -0.120 & -0.120 \\
         & H(\ce{CH2-$\alpha$})     & 2 &  \phantom{-}0.060 &  \phantom{-}0.120 \\
         & C(\ce{CO})               & 1 &  \phantom{-}0.470 &  \phantom{-}0.470 \\
         & O(\ce{CO})               & 1 & -0.470 & -0.470 \\
\hline
Ester    & C(\ce{CH2-$\alpha$})     & 1 &  \phantom{-}0.190 &  \phantom{-}0.190 \\
         & H(\ce{CH2-$\alpha$})     & 2 &  \phantom{-}0.030 &  \phantom{-}0.060 \\
         & C(\ce{C=O})              & 1 &  \phantom{-}0.490 &  \phantom{-}0.490 \\
         & O(\ce{C=O})              & 1 & -0.410 & -0.410 \\
         & O(\ce{-OR})              & 1 & -0.330 & -0.330 \\
\hline
\end{tabular}
\end{table}

\subsubsection{Crystallinity Metric}

To quantify the local structural ordering induced by the defects and thereby estimate the degree of crystallinity, a local bond-orientational order parameter was employed. For each backbone carbon atom, a unit orientation vector was defined by connecting the midpoints of its two adjacent backbone bonds. The orientational correlation between a reference vector and a neighbouring vector was quantified using the second Legendre polynomial,  \cite{Yamamoto2013JCP}

\begin{equation}
  P_2(\cos\theta)=\frac{3\cos^2\theta-1}{2},
  \label{eq:p2_crystallinity}
\end{equation}

where \(\theta\) is the angle between the two orientation vectors. The local order parameter associated with carbon atom \(i\), denoted \(\bar P_{2,i}\), was obtained by averaging \(P_2\) over all backbone carbons located within a cutoff radius of 0.7~nm. A carbon atom was classified as belonging to the crystalline phase when its local orientational order exceeded the threshold value \(P_{2,c}=0.6\). The overall crystallinity was then evaluated as the fraction of backbone carbon atoms identified as crystalline.

\subsubsection{Dielectric Properties from MD}
\label{sec:sc_methd}

The temperature dependent static dielectric constant, $\epss$, was computed from equilibrium fluctuations of the total dipole moment $\bm{M}$ via the Neumann fluctuation formula \cite{neumann1983}:
\begin{equation}
  \epss = \epsinf + \frac{\langle |\bm{M}|^2 \rangle
    - |\langle \bm{M} \rangle|^2}{3\epsilon_0 V k_\mathrm{B} T},
  \label{eq:eps0_fluct}
\end{equation}
where $V$ is the simulation-cell volume, $T$ is the temperature, and $\epsilon_0$ is the vacuum permittivity. The high-frequency dielectric constant $\epsinf$ used here was taken from the DFPT calculations. To analyze the dipolar relaxation dynamics, we computed the normalized autocorrelation function (ACF) of the total dipole moment,
\begin{equation}
  \phi(t) = \frac{\langle \bm{M}(0) \cdot \bm{M}(t) \rangle}
    {\langle |\bm{M}(0)|^2 \rangle},
  \label{eq:acf}
\end{equation}
which describes the decay of the initial dipolar orientation with time. To extract characteristic relaxation times, $\phi(t)$ was fitted to a Kohlrausch--Williams--Watts (KWW) stretched-exponential function \cite{williams1970,kohlrausch1854}:
\begin{equation}
  \phi(t) = A \exp\!\left[-\left(\frac{t}{\tau}\right)^{\!\alpha}\right],
  \label{eq:kww}
\end{equation}
where $\tau$ is the characteristic relaxation time, $\alpha$ ($0 < \alpha \le 1$) is the stretching exponent describing the breadth of the relaxation-time distribution, and $A$ is a prefactor that accounts for any fast initial decay. The complex frequency-dependent dielectric permittivity, $\eps^{*}(\omega)=\eps'(\omega)-i\epspp(\omega)$, was then obtained from the Fourier--Laplace transform of the KWW-fitted autocorrelation function:
\begin{equation}
  \eps^{*}(\omega)-\epsinf
  =
  \left[\epss-\epsinf\right]
  \int_{0}^{\infty}
  \left(-\frac{d\phi(t)}{dt}\right)
  e^{-i\omega t}\,dt.
  \label{eq:eps_complex}
\end{equation}
Accordingly, the dielectric loss spectrum, $\epspp(\omega)$, can be written as
\begin{equation}
  \epspp(\omega)
  =
  \left[\epss-\epsinf\right]
  \omega
  \int_{0}^{\infty}
  \phi(t)\cos(\omega t)\,dt.
  \label{eq:eps_loss}
\end{equation}

To quantify the contribution of each dipole component to the total relaxation, 
the total dipole moment was decomposed as $\Mtot = \Mdef + \Mmat$ and the raw (i.e., unnormalized) ACF was  partitioned into self and cross terms: $C_\mathrm{tot}(t) = C_\mathrm{self,def}(t) + C_\mathrm{self,mat}(t) + C_\mathrm{cross}(t)$\cite{henot2023orientational} \cite{sarkar2026calculation} (where $\mathrm{mat}\equiv \mathrm{matrix}$ and $\mathrm{def}\equiv \mathrm{defect}$).
Accordingly, the total ACF can be written as a sum of self and cross contributions
\begin{align}
C_{\mathrm{tot}}(t)
&= \langle \mathbf{M}_{\mathrm{def}}(0) \cdot \mathbf{M}_{\mathrm{def}}(t) \rangle
 + \langle \mathbf{M}_{\mathrm{mat}}(0) \cdot \mathbf{M}_{\mathrm{mat}}(t) \rangle \nonumber \\
&\quad + \langle \mathbf{M}_{\mathrm{def}}(0) \cdot \mathbf{M}_{\mathrm{mat}}(t) \rangle
 + \langle \mathbf{M}_{\mathrm{mat}}(0) \cdot \mathbf{M}_{\mathrm{def}}(t) \rangle.
\label{eq:Cexpanded}
\end{align}
Each component was independently normalized and then combined via dielectric weight 
fractions $w_i = C_i(0)/C_\mathrm{tot}(0) = \Deps_i/\Deps_\mathrm{total}$, so that the 
total normalized ACF is recovered as a weighted sum:
\begin{equation}
  \phi_\mathrm{tot}(t)
    = w_\mathrm{def}\,\phi_\mathrm{def}(t)
    + w_\mathrm{mat}\,\phi_\mathrm{mat}(t)
    + w_\mathrm{cross}\,\phi_\mathrm{cross}(t),
  \label{eq:weighted_acf}
\end{equation}
where $\phi_i(t) = C_i(t)/C_i(0)$ is the normalized ACF of each component. The weight 
fractions $w_i$ sum to unity and can be negative when the cross-correlation 
$\langle \delta\Mdef \cdot \delta\Mmat \rangle < 0$. The weighted representation 
$w_i \cdot \phi_i(t)$ directly shows the time-resolved contribution of each component to 
the total dielectric relaxation.

\section{Results and Discussion}
\label{sec:results}

\subsection{Morphology-Dependent Dielectric Response of Pristine Polyethylene}
\label{sec:morphology_pristine}

\begin{figure}[t]
    \centering
    \begin{tikzpicture}
        \node[anchor=south west, inner sep=0] (img) at (0,0) {
            \includegraphics[width=0.9\textwidth, trim={0 16cm 0 12cm}, clip]{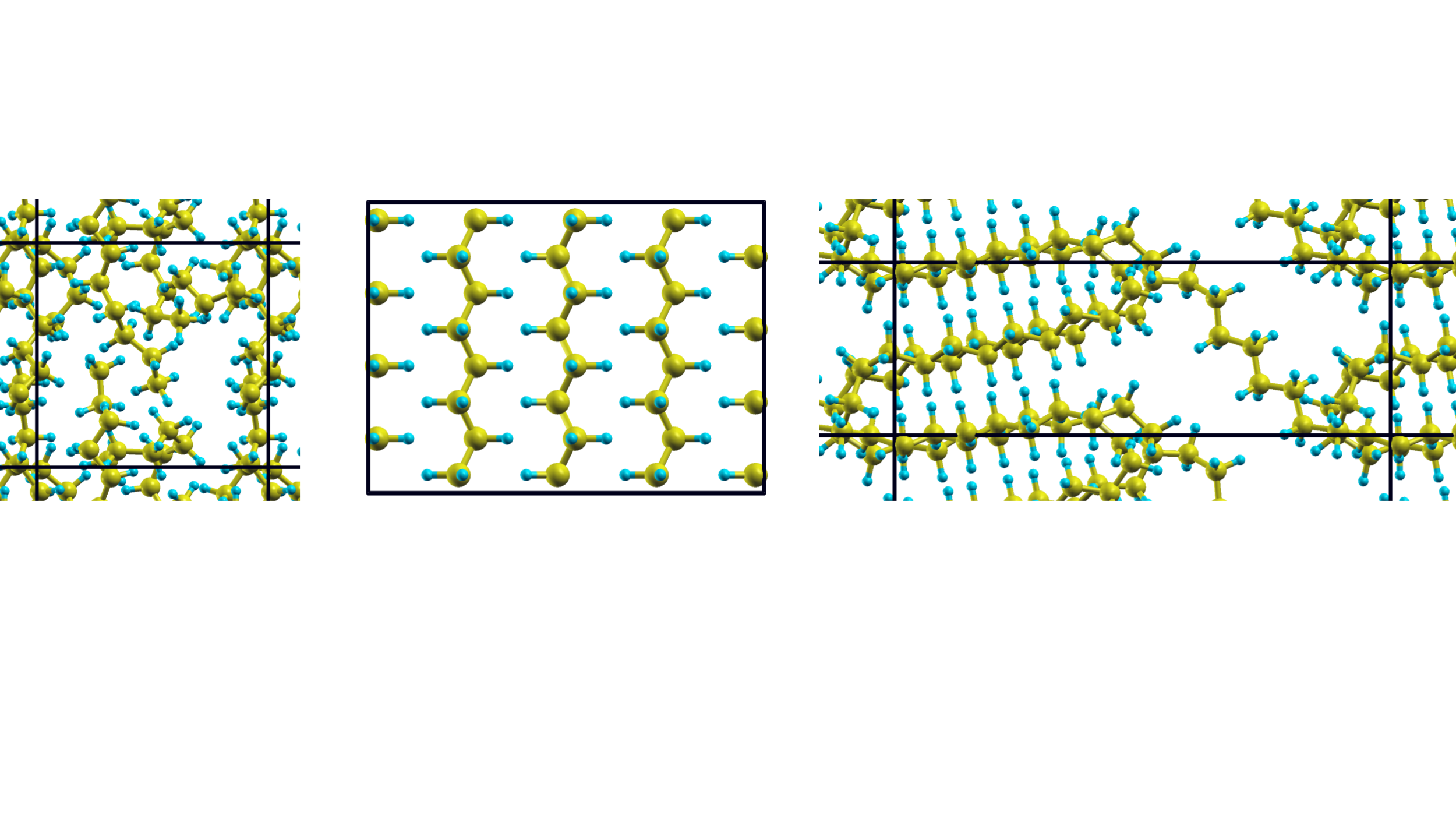}
        };

        \begin{scope}[shift={(0, 0)}] 

            \tikzset{
                axis/.style={->, >=stealth, line width=1.5pt, color=black}, %
                label/.style={font=\bfseries\small} %
            }

            \draw[axis] (0,0) -- (0.8, 0) node[label, right] {z}; 

        \end{scope}
    \end{tikzpicture}

\caption[Snapshots of the equilibrated systems]{
Snapshots of the equilibrated systems representing different morphological regions:  (Left) The amorphous domain with random chain orientation,  (Middle) the highly ordered crystalline lamell\ae, and (Right) the semi-crystalline interphase region. Yellow spheres represent carbon atoms, while blue spheres represent hydrogen.
}
\end{figure}

To isolate the influence of morphology on the dielectric response of pristine polyethylene, DFPT calculations were performed for three representative structural models: fully crystalline, fully amorphous, and semi-crystalline. Because polyethylene is intrinsically a semi-crystalline polymer and its dielectric behaviour is strongly influenced by chain packing and structural order, comparison across these idealised morphologies provides a useful framework for assessing how morphology governs both the electronic (high-frequency) permittivity, $\epsinf$, and the total static permittivity, $\epszero$. The corresponding dielectric constants are summarised in Table~\ref{tab:morphology}. 

The crystalline model, characterised by orthorhombic packing with perfectly aligned all-\emph{trans} chains, exhibits the highest average values: $\epsinf \approx 2.70$ and $\epszero \approx 2.73$. The elevated permittivity in the crystalline phase is primarily attributable to its higher density (1.08 $\text{g/cm}^3$) and more ordered molecular packing, which increase the number of polarisable electrons per unit volume and enhance the collective electronic response. The DFPT dielectric tensor for the crystalline phase shows pronounced anisotropy, consistent with the strong structural directionality of an aligned-chain polymer crystal. Specifically, $\epsinf$ along the chain axis ($\eps_z \approx 2.90$) is substantially larger than the in-plane components ($\eps_x, \eps_y \approx 2.56$), and a similar anisotropy persists in $\epszero$ ($\eps_z \approx 2.92$ vs.\ $\eps_{x,y} \approx 2.61$).

The fully amorphous model, comprising four decane-length chains within the simulation cell, yields $\epsinf \approx 2.38$ and $\epszero \approx 2.44$. In contrast to the crystalline case, the amorphous model is effectively isotropic, the three tensor components $\eps_x$, $\eps_y$, and $\eps_z$ agree to within 1--2\% of each other, because the randomly oriented chains average out any single-chain anisotropy across all directions.

The semi-crystalline model, which incorporates both ordered (crystal-like) and disordered (amorphous-like) regions separated by interfacial surfaces, provides the most realistic representation of typical PE morphology. This model yields $\epsinf \approx 2.23$ and $\epszero \approx 2.26$, averaged over all tensor components. The computed tensor components ($\eps_x$, $\eps_y$, $\eps_z$) differ by approximately 1\% or less, indicating that the semi-crystalline model is effectively quasi-isotropic. The resulting effective permittivity of $\sim$2.25 closely matches the bulk experimental values for common PE, typically reported as $2.3 \pm 0.1$ \cite{bur1985dielectric}, thereby underscoring the validity of this mixed-phase model.

These findings highlight the advantage of the semi-crystalline model in capturing the dielectric response of real polyethylene, as it provides a more realistic representation of the dielectric behaviour of practical PE systems than either the fully amorphous or fully crystalline limit. While crystalline domains exhibit pronounced anisotropic polarization at the local scale, their coexistence with amorphous regions leads to an overall response that is effectively isotropic at the macroscopic level. Accordingly, all subsequent DFPT calculations reported in this work are based on the semi-crystalline slab model, ensuring that both the baseline dielectric response and the defect-induced variations are assessed in a structurally representative PE environment.

\begin{table}[htbp]
  \centering
  \caption{DFPT-computed dielectric constants of pristine polyethylene
    for three models. The anisotropy ratio
    $\eps_z/\eps_{x,y}$ quantifies the difference in dielectric response along the polymer chain vs perpendicular to it.}
  \label{tab:morphology}
  \begin{tabular}{l cccc}
    \toprule
    {Morphology}
      & {$\epsinf$}
      & {$\epszero$}
      & {$\Delta\eps$}
      & {$\eps_z/\eps_{x,y}$} \\
    \midrule
    Crystalline      & 2.70 & 2.73 & 0.03 & $\sim$1.13 \\
    Amorphous        & 2.38 & 2.44 & 0.06 & $\sim$1.00 \\
    Semi-crystalline & 2.23 & 2.26 & 0.03 & $\sim$1.01 \\
    \addlinespace
    Experiment \cite{bur1985dielectric}
      & \multicolumn{2}{c}{$2.3 \pm 0.1$}
      & {---}
      & {---} \\
    \bottomrule
  \end{tabular}
\end{table}

\subsection{DFPT Results: Defect-Induced Dielectric Modifications}
\label{sec:dfpt_results}

\subsubsection{High-Frequency Dielectric Constant}
\label{sec:epsinf}

Figure~\ref{fig:dfpt_conc}a presents the DFPT-computed high-frequency dielectric constant, $\epsinf$, as a function of defect type and concentration, referenced against the pristine PE value of 2.232. In the high-frequency regime, the dielectric response is dominated by electronic polarization, and defect-induced changes in $\epsinf$ therefore directly probe how each functional group modifies the electronic polarizability of the PE matrix. \cite{Zha2021NanoEnergy,BakerJarvis2012JResNIST} These trends may be qualitatively interpreted in terms of the Clausius--Mossotti relation. \cite{BakerJarvis2012JResNIST}

The magnitude and sign of the shift in $\epsinf$ are specific to the chemical nature of the defect. Defects that enhance the net electronic polarizability of the matrix increase $\epsinf$, whereas those that reduce it decrease $\epsinf$ relative to pristine PE. The calculated trends separate into two groups. Enone, alcohol, ketone, and vinylene increase $\epsinf$ relative to pristine PE, whereas aldehyde, ester, and carboxylic-acid defects reduce it. The largest positive shifts are observed for alcohol and enone, for which $\epsinf$ increases to about 2.244--2.245 at 2~mol\,\% and to 2.254--2.257 at 4~mol\,\%. Ketone and vinylene produce smaller but still positive changes, reaching approximately 2.243 and 2.236 at 4~mol\,\%, respectively.

The increase in $\epsinf$ for enone and vinylene is consistent with the enhanced electronic deformability associated with unsaturated and conjugated bonding motifs. \cite{Cho2008ProgPolymSci} Alcohol and ketone also lead to positive shifts, indicating that these defects increase the local electronic polarizability in the present semi-crystalline PE environment. In contrast, aldehyde, ester, and carboxylic-acid defects lower $\epsinf$ below the pristine value, with the strongest reductions found for ester and aldehyde, which reach 2.199 and 2.203, respectively, at 4~mol\,\%. These results indicate that, in the present systems, oxygen-rich carbonyl-containing defects do not translate into an increase in the net electronic polarizability of the matrix, in contrast to previous findings on a family of cyclic polyolefins. \cite{Wu2023JAdvDielect}

\subsubsection{Static Dielectric Constant}
\label{sec:eps0}

\begin{figure}[htbp]
  \centering
  \begin{subfigure}[t]{0.48\textwidth}
    \centering
    \includegraphics[width=\linewidth]{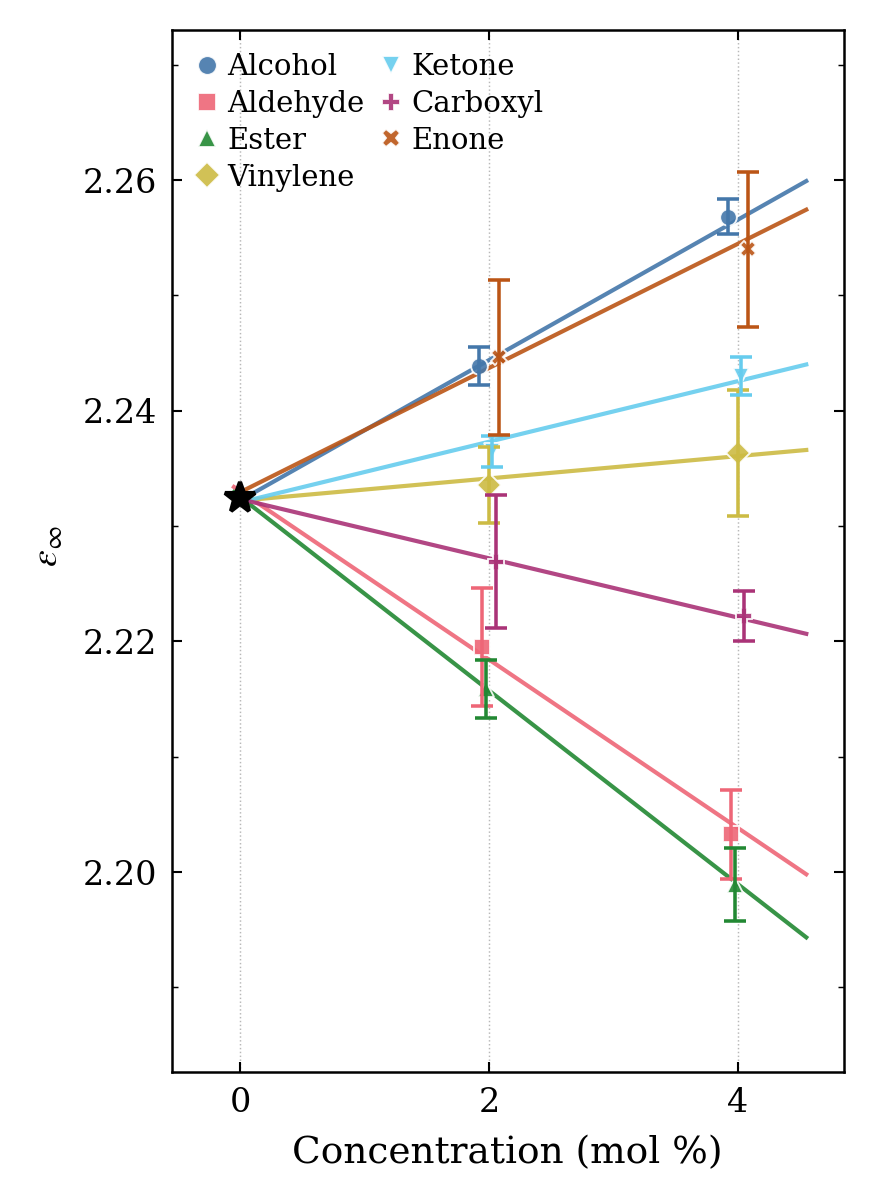}
    \caption{$\epsinf$ vs.\ concentration}
    \label{fig:dfpt_epsinf}
  \end{subfigure}\hfill
  \begin{subfigure}[t]{0.48\textwidth}
    \centering
    \includegraphics[width=\linewidth]{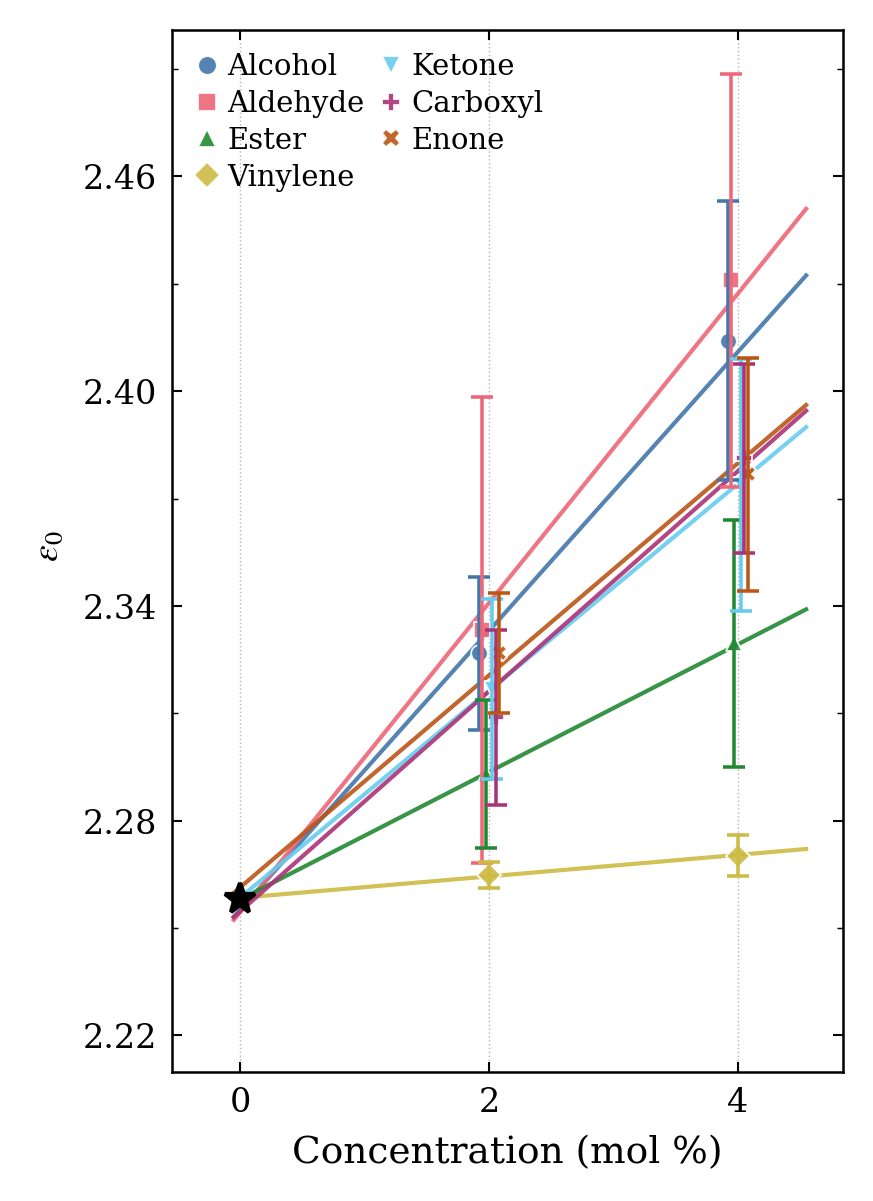}
    \caption{$\epszero$ vs.\ concentration}
    \label{fig:dfpt_eps0}
  \end{subfigure}
  \caption{DFPT-computed dielectric constants as a function of defect
    concentration for all seven defect types.
    (a)~High-frequency dielectric constant $\epsinf$ (electronic contribution only).
    (b)~Static dielectric constant $\epszero$ (electronic + ionic contributions).
    The black star denotes the pristine PE value. Error bars represent the
    configurational spread over different defect placements within the supercell.
    Solid lines are linear fits.}
  \label{fig:dfpt_conc}
\end{figure}

Figure~\ref{fig:dfpt_conc}b  presents the DFPT-computed static dielectric constant $\epszero$ as a function of defect type and concentration, referenced against the pristine PE value of 2.258. The error bars represent the configurational spread obtained from symmetry-inequivalent defect placements in the supercell, and the solid lines are linear fits used to highlight the concentration dependence. In contrast to $\epsinf$, which contains only the electronic contribution, $\epszero$ additionally includes the lattice contribution arising from $\Gamma$-point IR-active phonon modes. Within DFPT, the static dielectric tensor is calculated using Equation~\eqref{eq:eq2_mode_resolved} where the lattice contribution relies on both dynamical charge and phonon frequency, such that low-frequency polar modes make the largest contributions to $\epszero$. \cite{baroni2001,Gonze1997PRB}

Consistent with this framework, all seven defect species increase $\epszero$ above the pristine value at both concentrations, in clear contrast to the mixed behavior observed for $\epsinf$. This indicates that each defect introduces additional lattice-level polarization channels, even when its effect on the purely electronic contribution is weak or negative. 
At 2~mol\,\%, most defect-induced shifts in $\epszero$ remain within overlapping error bars, so distinctions among individual defects should be interpreted cautiously. 
At 4~mol\,\%, the mean values separate more clearly and reveal a chemically specific ranking of the static dielectric response.

The numerical trends are summarized in Table~\ref{tab:dfpt_summary}. At 4~mol\,\%, aldehyde yields the largest static dielectric constant, $\epszero=2.431$, followed by
alcohol ($2.414$), carboxyl ($2.381$), enone ($2.377$), ketone ($2.374$), ester ($2.330$), and vinylene ($2.270$). When the ionic increment is isolated through $\Delta\eps = \epszero - \epsinf$, aldehyde again shows the strongest ionic contribution
($\Delta\eps = 0.228$), whereas vinylene remains close to the pristine reference value of 0.026, reaching only $0.034$ at 4~mol\,\%. These results demonstrate that the static dielectric response is substantially more sensitive to defect chemistry than the
high-frequency dielectric response.

The trends in Figure~\ref{fig:dfpt_conc}b and Table~\ref{tab:dfpt_summary} also show that the ranking of $\epszero$ cannot be inferred from $\epsinf$ alone. For example, alcohol and enone both increase $\epsinf$, but alcohol produces a much larger increase in $\epszero$ because its ionic contribution is also substantial. Conversely, aldehyde lowers $\epsinf$ relative to pristine PE while still producing the largest increase in
$\epszero$. This contrast shows that the static dielectric response is governed by the combined influence of the electronic background and defect-activated lattice polarization.

Aldehyde provides the clearest illustration of this interplay. Despite having the smallest carbonyl-related Born effective charges among the oxygen-containing carbonyl defects considered, it exhibits the largest ionic increment. This indicates that the large increase in $\epszero$ cannot be attributed to Born effective charge magnitude
alone and is instead consistent with a particularly strong contribution from low-frequency IR-active modes through the $1/\omega_m^2$ weighting in Eq.~\eqref{eq:eq2_mode_resolved}. Ester provides the complementary case: although it carries the largest carbonyl-related Born effective charges, its ionic increment is more moderate ($\Delta\eps = 0.131$), consistent with a smaller contribution from low-frequency polar modes.

Carboxyl, ketone, and enone form an intermediate group at 4~mol\,\%, with $\epszero$ values between 2.374 and 2.381 and ionic increments of approximately 0.123--0.159. Alcohol belongs to the more strongly enhancing class, suggesting that hydroxyl-containing defects can also couple efficiently to IR-active lattice modes. Vinylene stands apart
from all other defects: its static dielectric constant remains only slightly above that of pristine PE, and its ionic increment is the smallest in the series. This weak response is consistent with the near-symmetric nature of the \ce{C=C} unit, which carries only small dynamical charges and therefore contributes little to the IR-active lattice polarization.

\begin{table}[htbp]
  \centering
  \caption{DFPT-computed dielectric constants for pristine and defective-PE. $\epsinf$: high-frequency (electronic) dielectric constant;
    $\epszero$: static dielectric constant (electronic + ionic);
    $\Delta\eps = \epszero - \epsinf$: ionic contribution.
    Defects are ordered by decreasing $\epszero$ at 4~mol\,\%.}
  \label{tab:dfpt_summary}
  \sisetup{round-mode=places, round-precision=2}
  \begin{tabular}{l *{3}{S[table-format=1.3]} *{3}{S[table-format=1.3]}}
    \toprule
    & \multicolumn{3}{c}{2~mol\,\%}
    & \multicolumn{3}{c}{4~mol\,\%} \\
    \cmidrule(lr){2-4} \cmidrule(lr){5-7}
    {Defect}
      & {$\epsinf$} & {$\epszero$} & {$\Delta\eps$}
      & {$\epsinf$} & {$\epszero$} & {$\Delta\eps$} \\
    \midrule
    Pristine PE
      & 2.23 & 2.26 & 0.03
      & {---}  & {---}  & {---} \\
    \addlinespace
    Aldehyde
      & 2.22 & 2.33 & 0.11
      & 2.20 & 2.43 & 0.23 \\
    Alcohol
      & 2.24 & 2.33 & 0.08
      & 2.26 & 2.42 & 0.16 \\
    Carboxyl
      & 2.23 & 2.31 & 0.08
      & 2.22 & 2.38 & 0.16 \\
    Enone
      & 2.25 & 2.33 & 0.08
      & 2.26 & 2.38 & 0.12 \\
    Ketone
      & 2.24 & 2.32 & 0.08
      & 2.24 & 2.37 & 0.13 \\
    Ester
      & 2.22 & 2.29 & 0.07
      & 2.20 & 2.33 & 0.13 \\
    Vinylene
      & 2.23 & 2.26 & 0.03
      & 2.24 & 2.27 & 0.03 \\
    \bottomrule
  \end{tabular}
\end{table}

\subsubsection{Born Effective Charges}
\label{sec:bec}

Table~\ref{tab:bec} reports the isotropic Born effective charges, $Z^* = \mathrm{Tr}(\mathbf{Z}^*)/3$, of the chemically active defect atoms for each defect-doped PE system at 4~mol\,\%, computed from DFPT with the acoustic sum rule (ASR) applied. Born effective charges measure the change in macroscopic polarization induced by atomic displacement and therefore quantify how strongly a given atom can contribute to IR-active lattice polarization \cite{baroni2001,Gonze1997PRB}. For reference, the pristine PE backbone exhibits only small dynamical charges, with $Z^*_{\mathrm{C}} = +0.101 \pm 0.028$ and
$Z^*_{\mathrm{H}} = -0.052 \pm 0.016$, consistent with the nearly non-polar character of the \ce{-CH2-} repeat unit.

The defect atoms display much larger and chemically specific Born effective charges than the pristine backbone atoms. Among the carbonyl-containing defects, the carbonyl carbon charges increase in the order aldehyde ($+0.947$), ketone ($+1.039$), enone ($+1.223$),
carboxyl ($+1.370$), and ester ($+1.412$), while the corresponding carbonyl oxygens carry large negative values between $-0.875$ and $-1.079$. These values confirm that the \ce{C=O} motif is a strong source of dynamical charge in defected PE. Likewise, hydroxyl-bearing defects show substantial dynamical charges on both oxygen and hydrogen, with $Z^*(\mathrm{O_{OH}})=-0.842$ and $Z^*(\mathrm{H_{pol}})=+0.326$ for alcohol, and $Z^*(\mathrm{O_{OH}})=-0.914$ and $Z^*(\mathrm{H_{pol}})=+0.382$ for carboxyl. These values are far larger in magnitude than those of the pristine backbone and help explain why hydroxyl-containing defects produce sizeable ionic contributions to $\epszero$.

At the same time, Table~\ref{tab:bec}, when compared to Table~\ref{tab:dfpt_summary}, makes clear that large Born effective charges do not by themselves determine the final ranking of $\Delta\eps$. Ester has the largest carbonyl-related dynamical charges in the series, yet its static dielectric enhancement is smaller than that of aldehyde. Conversely, aldehyde exhibits the largest ionic increment while carrying the smallest carbonyl-related Born effective charges among the oxygen-containing carbonyl defects. This comparison reinforces the conclusion drawn from Eq.~\eqref{eq:eq2_mode_resolved}: the lattice contribution is controlled by the interplay between dynamical charge and phonon frequency. Large $Z^*$ values are necessary for
strong IR activity, but they lead to the largest dielectric response only when coupled to sufficiently soft polar modes.

The enone defect is also instructive because the dynamical charge is not confined to the carbonyl pair alone. In addition to the carbonyl carbon and oxygen, the adjacent sp$^2$ carbon carries a large positive Born effective charge of $+1.132 \pm 0.074$, showing that the conjugated defect redistributes dynamical charge over multiple atoms. This delocalized polar response is consistent with the intermediate ionic contribution of enone. By contrast, vinylene exhibits a nearly negligible isotropic Born effective charge on the sp$^2$ carbons ($-0.077 \pm 0.028$), only slightly different from the pristine
backbone value. Its weak static dielectric enhancement is therefore fully consistent with the absence of a strongly polar vibrational channel.

A mode-resolved decomposition of the lattice contribution to $\epszero$ is performed to analyze the influence by low-frequency IR-active modes in determining the ionic increment. For each defect configuration at 4~mol\,\%, the positive dielectric mode weight $\delta\eps_m$ of every IR-active phonon mode was extracted and assigned to frequency bins. Table~\ref{tab:mode_fractions} summarizes the key numerical results averaged over the five symmetry-inequivalent configurations per defect type. These data provide a quantitative, mode-resolved explanation for the defect-specific ranking of $\Delta\eps$.  This trend is most pronounced for aldehyde, for which 68.5\% of the cumulative dielectric mode weight lies below 100~cm$^{-1}$ and 80.4\% below 200~cm$^{-1}$, together with the lowest dominant-mode frequency, $\omega_{\mathrm{top}} = 40.1 \pm 20.2$~cm$^{-1}$. In contrast, alcohol, despite its relatively large $\Deps$, shows, on average, much smaller low-frequency fractions (20.8\% below 100~cm$^{-1}$ and 37.3\% below 200~cm$^{-1}$) and, apart from one specific locations, high dominant-mode frequency above 250~cm$^\text{-1}$,
indicating that its dielectric enhancement is distributed over higher-frequency IR-active modes.  The remaining defects exhibit intermediate behavior, with roughly 41--53\% of the dielectric mode weight below 100~cm$^\text{-1}$ and 51--65\% below 200~cm$^\text{-1}$. Overall, these results show that the defect-induced increase in the static dielectric constant cannot be inferred from defect chemistry alone, but instead reflects the combined influence of electronic structure, dynamical charge, and the frequency distribution of IR-active phonon modes.

\begin{table}[htbp]
  \centering
  \small
  \setlength{\tabcolsep}{4pt}
  \caption{Isotropic Born effective charges $Z^* = \mathrm{Tr}(\mathbf{Z}^*)/3$
    (ASR-corrected) of the chemically active defect atoms at 4~mol\,\%,
    from DFPT. Values are mean $\pm$ std over symmetry-inequivalent configurations.
    `--': role chemically absent.
    $^{a}$Enone: $\mathrm{C}_\alpha$ and $\mathrm{C}_\beta$ from instance~1 only.
    $^{b}$Vinylene: average over both \ce{C=C} carbons.}
  \label{tab:bec}
  \begin{tabular}{l cccccc}
    \toprule
    & \multicolumn{2}{c}{Carbon}
    & \multicolumn{3}{c}{Oxygen}
    & {Hydrogens} \\
    \cmidrule(lr){2-3}\cmidrule(lr){4-6}\cmidrule(lr){7-7}
	  pristine PE & \multicolumn{2}{c}{$0.10\pm0.03$} &  \multicolumn{3}{c}{--} & $-0.05\pm0.02$ \\ \midrule
    {Defect}
      & {$Z^*(\mathrm{C_{C\!=\!O}})$}
      & {$Z^*(\mathrm{C_{sp^2}})$}
      & {$Z^*(\mathrm{O_{C\!=\!O}})$}
      & {$Z^*(\mathrm{O_{br}})$}
      & {$Z^*(\mathrm{O_{OH}})$}
      & {$Z^*(\mathrm{H_{pol}})$} \\
    \midrule
    Alcohol
      & -- & --
      & -- & --
      & $-0.84\pm0.03$
      & $+0.33\pm0.02$ \\
    Aldehyde
      & $+0.95\pm0.02$ & --
      & $-0.875\pm0.02$ & -- & --
      & $+0.05\pm0.01$ \\
    Ketone
      & $+1.04\pm0.06$ & --
      & $-0.91\pm0.03$ & -- & -- & -- \\
    Ester
      & $+1.41\pm0.06$ & --
      & $-1.08\pm0.09$
      & $-0.99\pm0.09$
      & -- & -- \\
    Carboxyl
      & $+1.37\pm0.03$ & --
      & $-0.95\pm0.06$ & --
      & $-0.91\pm0.06$
      & $+0.38\pm0.02$ \\
    Enone$^{a}$
      & $+1.22\pm0.06$
      & $+1.13\pm0.07$
      & $-0.96\pm0.04$
      & -- & -- & -- \\
    Vinylene$^{b}$
      & --
      & $-0.08\pm0.03$
      & -- & -- & -- & -- \\
    \bottomrule
  \end{tabular}
\end{table}

\begin{table*}[htbp]
  \centering
  \caption{Mode-resolved dielectric weight analysis at 4~mol\,\%.
    $\Delta\eps$: ionic increment ($\epszero - \epsinf$);
    $f_{\le 100}$, $f_{\le 200}$, $f_{\le 400}$: cumulative fraction
    of positive dielectric mode weight from modes below the indicated
    frequency threshold (in cm$^{-1}$);
    dominant frequency bin: the single bin carrying the largest weight fraction.
    All values are averages $\pm$ std over five symmetry-inequivalent
    configurations.  Defects are ordered by decreasing $\Delta\eps$.}
  \label{tab:mode_fractions}
  \small
  \setlength{\tabcolsep}{5pt}
  \begin{tabular}{l S[table-format=1.3] @{${}\pm{}$} S[table-format=1.3]
                    S[table-format=1.2] @{${}\pm{}$} S[table-format=1.2]
                    S[table-format=1.2] @{${}\pm{}$} S[table-format=1.2]
                    S[table-format=1.2] @{${}\pm{}$} S[table-format=1.2]
                    l}
    \toprule
    {Defect}
      & \multicolumn{2}{c}{$\Delta\eps$}
      & \multicolumn{2}{c}{$f_{\le 100}$}
      & \multicolumn{2}{c}{$f_{\le 200}$}
      & \multicolumn{2}{c}{$f_{\le 400}$}
      & {Dominant bin} \\
    \midrule
    Aldehyde
      & 0.218 & 0.075
      & 0.69  & 0.08
      & 0.80  & 0.06
      & 0.84  & 0.03
      & (0, 100] \\
    Alcohol
      & 0.179 & 0.088
      & 0.21  & 0.10
      & 0.37  & 0.22
      & 0.80  & 0.07
      & (200, 400] \\
    Carboxyl
      & 0.151 & 0.012
      & 0.45  & 0.09
      & 0.53  & 0.08
      & 0.55  & 0.08
      & (0, 100] \\
    Ketone
      & 0.125 & 0.038
      & 0.53  & 0.14
      & 0.65  & 0.08
      & 0.71  & 0.07
      & (0, 100] \\
    Ester
      & 0.130 & 0.038
      & 0.41  & 0.11
      & 0.51  & 0.08
      & 0.61  & 0.06
      & (0, 100] \\
    Enone
      & 0.120 & 0.016
      & 0.47  & 0.07
      & 0.62  & 0.06
      & 0.67  & 0.05
      & (0, 100] \\
    Vinylene
      & 0.034 & 0.002
      & 0.15  & 0.05
      & 0.24  & 0.04
      & 0.29  & 0.03
      & (1600, 4000] \\
    \bottomrule
  \end{tabular}
\end{table*}

\subsubsection{Position-Dependent Dielectric Response}
\label{sec:position}

\begin{figure}[htbp]
  \centering
  \begin{subfigure}[t]{\textwidth}
    \centering
    \includegraphics[width=\linewidth]{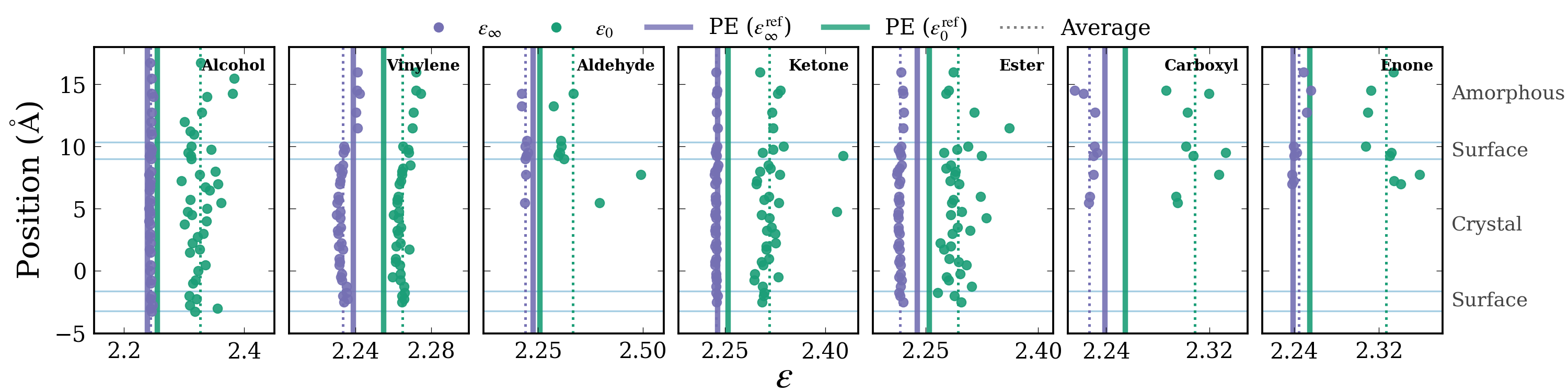}
    \caption{2~mol\,\%}
  \end{subfigure}\\[4pt]
  \begin{subfigure}[t]{\textwidth}
    \centering
    \includegraphics[width=\linewidth]{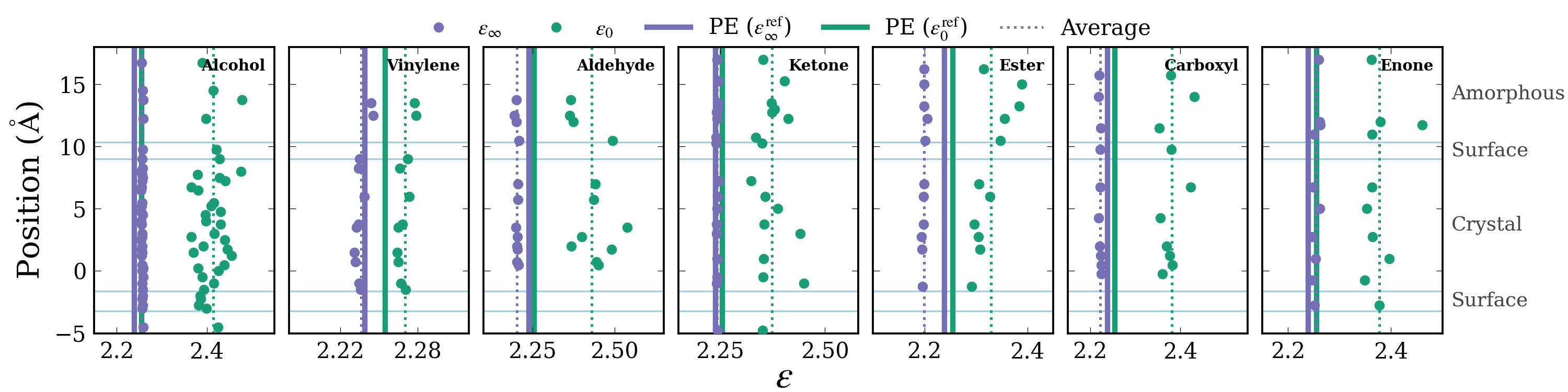}
    \caption{4~mol\,\%}
  \end{subfigure}
  \caption{Position-dependent DFPT dielectric response for all seven defect types at
    (a)~2~mol\,\% and (b)~4~mol\,\%.  Purple markers: $\epsinf$; green markers: $\epszero$.
    Vertical bands indicate the pristine PE reference values. Horizontal
    lines separate the crystalline core, surface/interfacial, and amorphous regions.
    The dotted vertical lines denote the configurational averages.}
  \label{fig:dfpt_pos}
\end{figure}

Figures~\ref{fig:dfpt_pos}a and \ref{fig:dfpt_pos}b show the position-dependent dielectric response for defects placed at different locations along the lamellar normal (crystalline, interfacial, and amorphous regions) at 2 and 4~mol\,\%, respectively. The purple and green markers represent $\epsinf$ and $\epszero$, while the vertical bands denote the pristine PE reference values. The horizontal blue lines delineate the crystalline, surface (interfacial), and amorphous regions of the slab model. Across all defect types and both concentrations, the $\varepsilon_\infty$ data points remain tightly clustered about the pristine PE reference, with a spread of approximately $0.03$--$0.08$ in the 2\,mol\% case and $0.03$--$0.12$ at 4\,mol\%. In contrast, the $\varepsilon_0$ points exhibit a substantially wider distribution at every position, with spreads ranging from ${\sim}0.04$ (vinylene, 2\,mol\%) to ${\sim}0.30$ (aldehyde, 4\,mol\%). Aldehyde displays the widest $\varepsilon_0$ distribution among all systems at both concentrations. At 2\,mol\%, individual configurations span from values close to the pristine $\varepsilon_0^{\rm ref}$ up to ${\sim}2.50$; at 4\,mol\%, the spread broadens further and extends to ${\sim}2.55$. The remaining carbonyl-bearing defects --- ketone, carboxyl, ester, and enone --- occupy an intermediate range of $\varepsilon_0$ dispersions. At 2\,mol\%, their $\varepsilon_0$ spreads lie in the range $0.08$--$0.16$, and at 4\,mol\%, the dispersions widen reaching $0.20$--$0.25$. In all cases, the $\varepsilon_0$ spread widens systematically from 2 to 4\,mol\%, consistent with the doubling of IR-active defect units.

\subsection{Molecular Dynamics Results}
\label{sec:md_results}
Classical molecular dynamics (MD) simulations were performed to analyze the time-dependent reorientation of the defect dipoles and the associated frequency-dependent permittivity and dielectric loss spectra. While DFPT calculations provide insights into the static and high-frequency dielectric response from electronic and ionic contributions, they do not capture the orientational polarization dynamics that contributes to the dielectric permittivity which dominates in the low-frequency regions at finite temperature. The MD simulations focus on four defects (alcohol, aldehyde, ester, and ketone) at concentrations of 2 and 4~mol\,\%.

\subsubsection{Morphology and Crystallinity}
\label{sec:crystallinity}

\begin{figure}[htbp]
  \centering
  \begin{subfigure}[t]{0.32\textwidth}
    \centering
    \includegraphics[width=\linewidth]{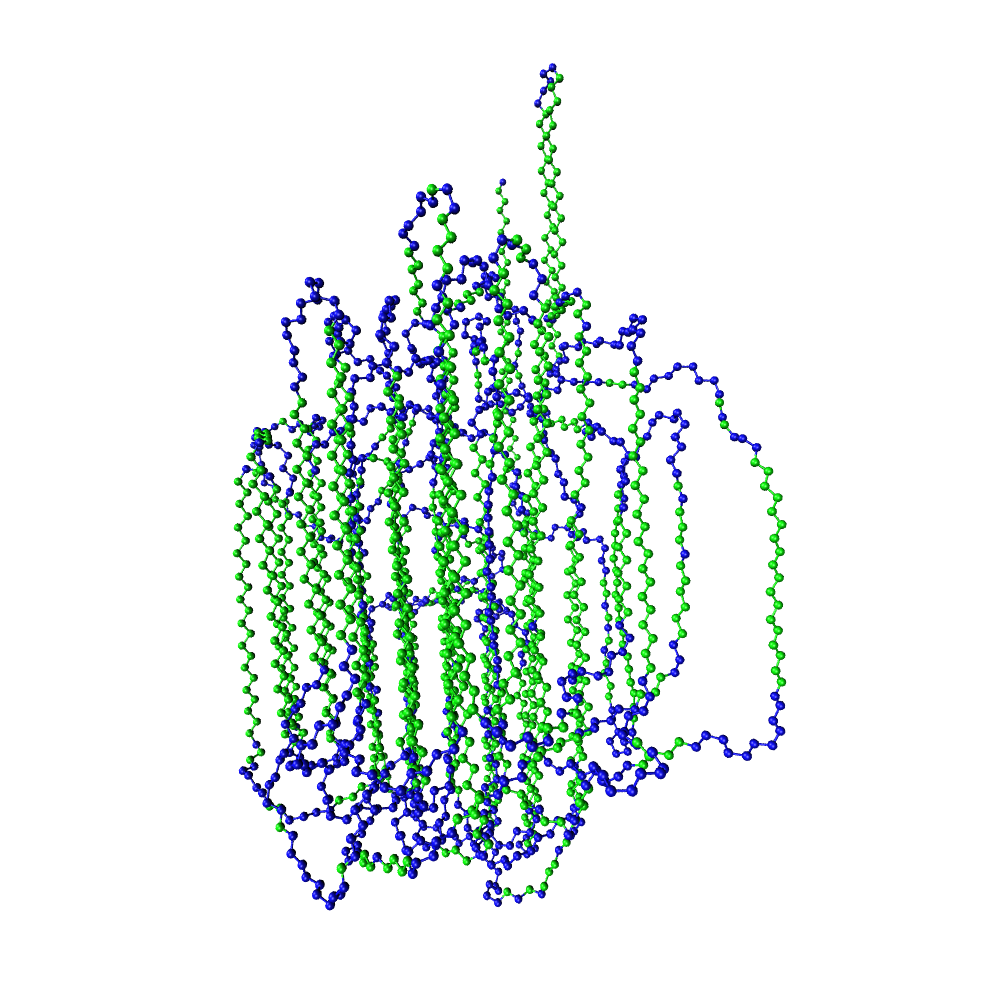}
    \caption{Pristine PE}
  \end{subfigure}\hfill
  \begin{subfigure}[t]{0.32\textwidth}
    \centering
    \includegraphics[width=\linewidth]{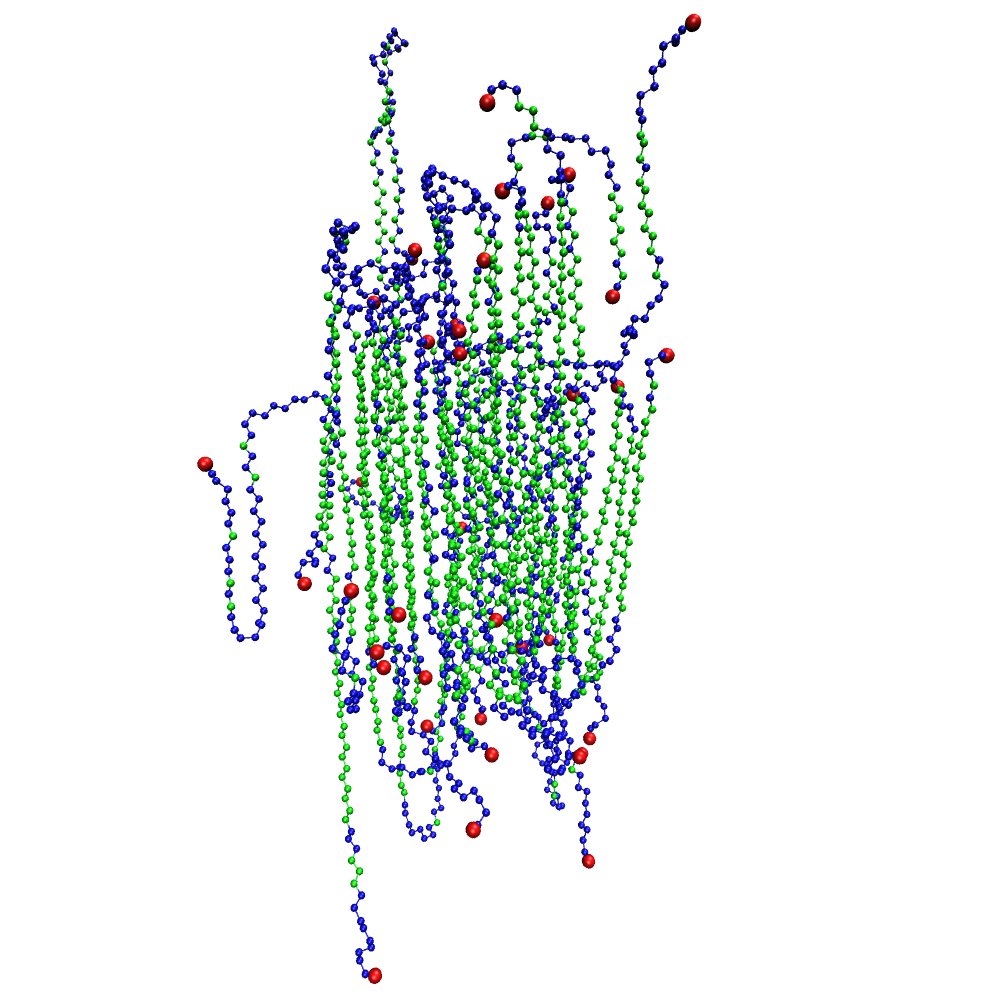}
    \caption{Aldehyde}
  \end{subfigure}\hfill
  \begin{subfigure}[t]{0.32\textwidth}
    \centering
    \includegraphics[width=\linewidth]{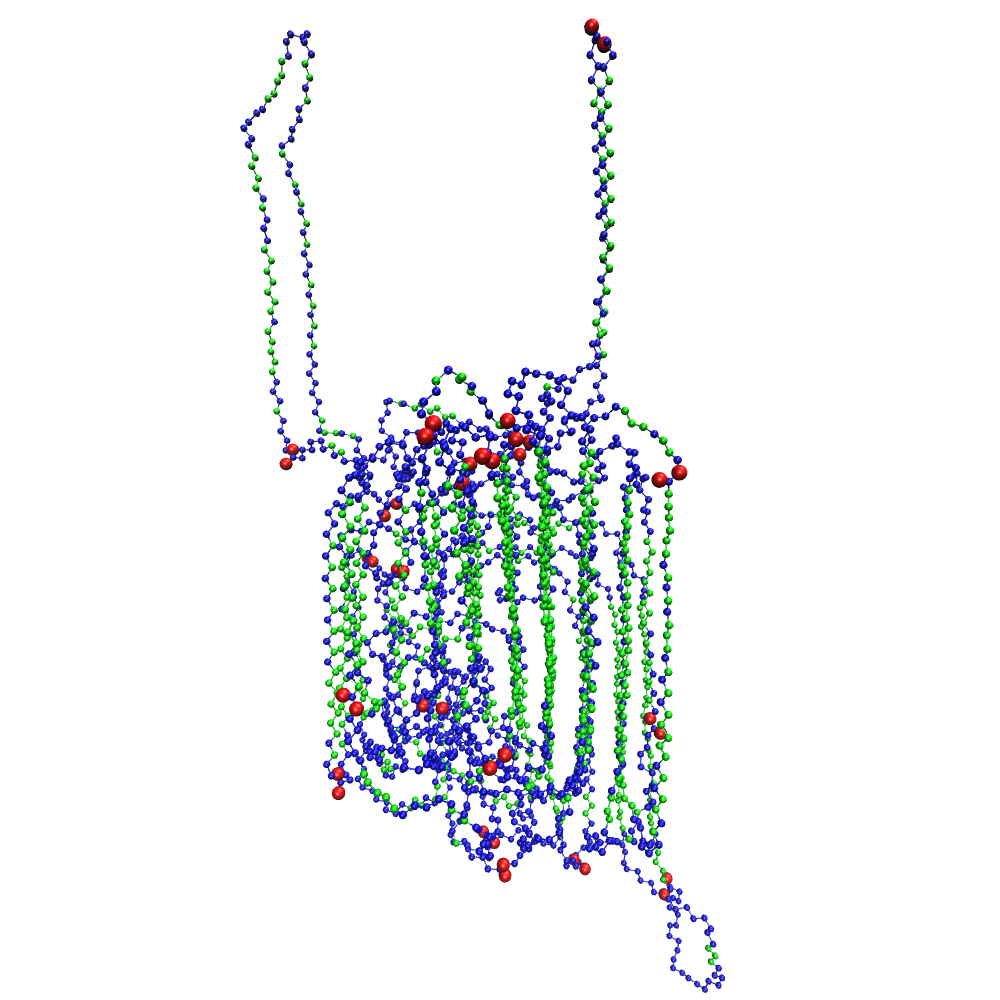}
    \caption{Ester}
  \end{subfigure}\\[6pt]
  \begin{subfigure}[t]{0.32\textwidth}
    \centering
    \includegraphics[width=\linewidth]{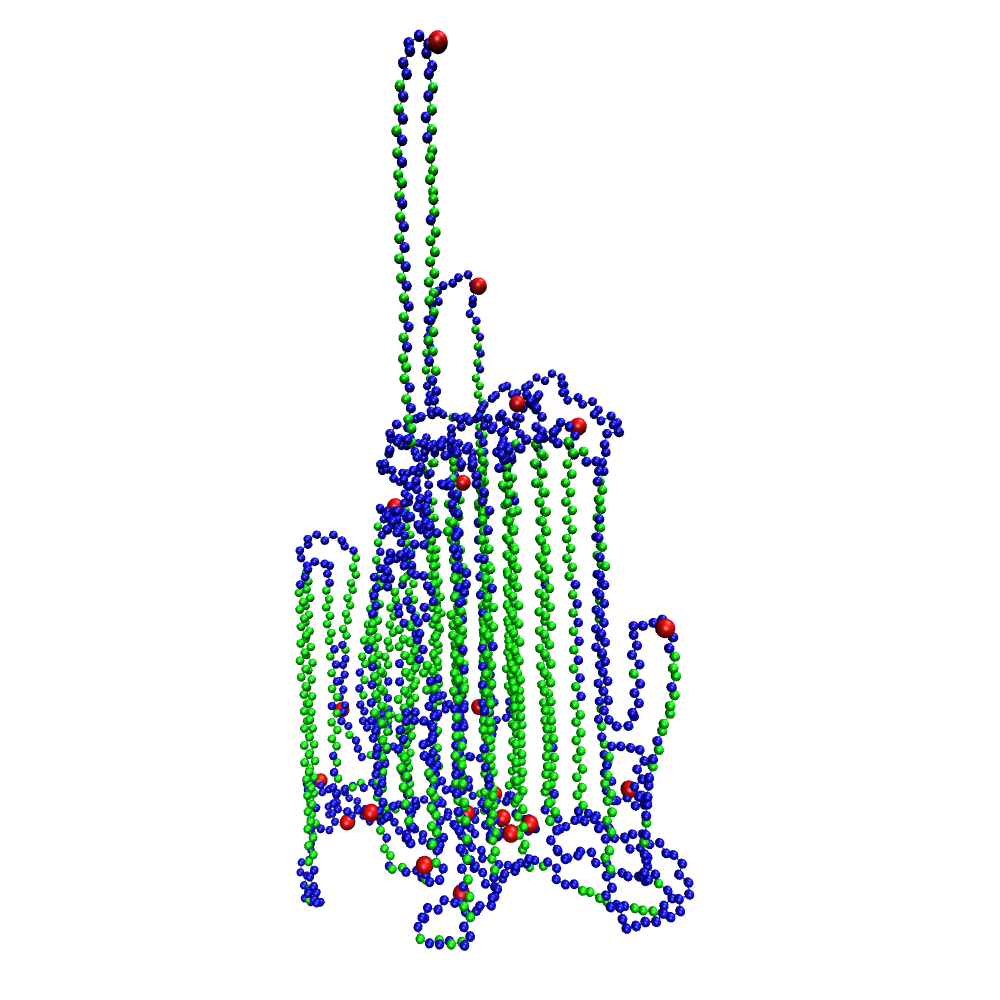}
    \caption{Ketone}
  \end{subfigure}\hfill
  \begin{subfigure}[t]{0.32\textwidth}
    \centering
    \includegraphics[width=\linewidth]{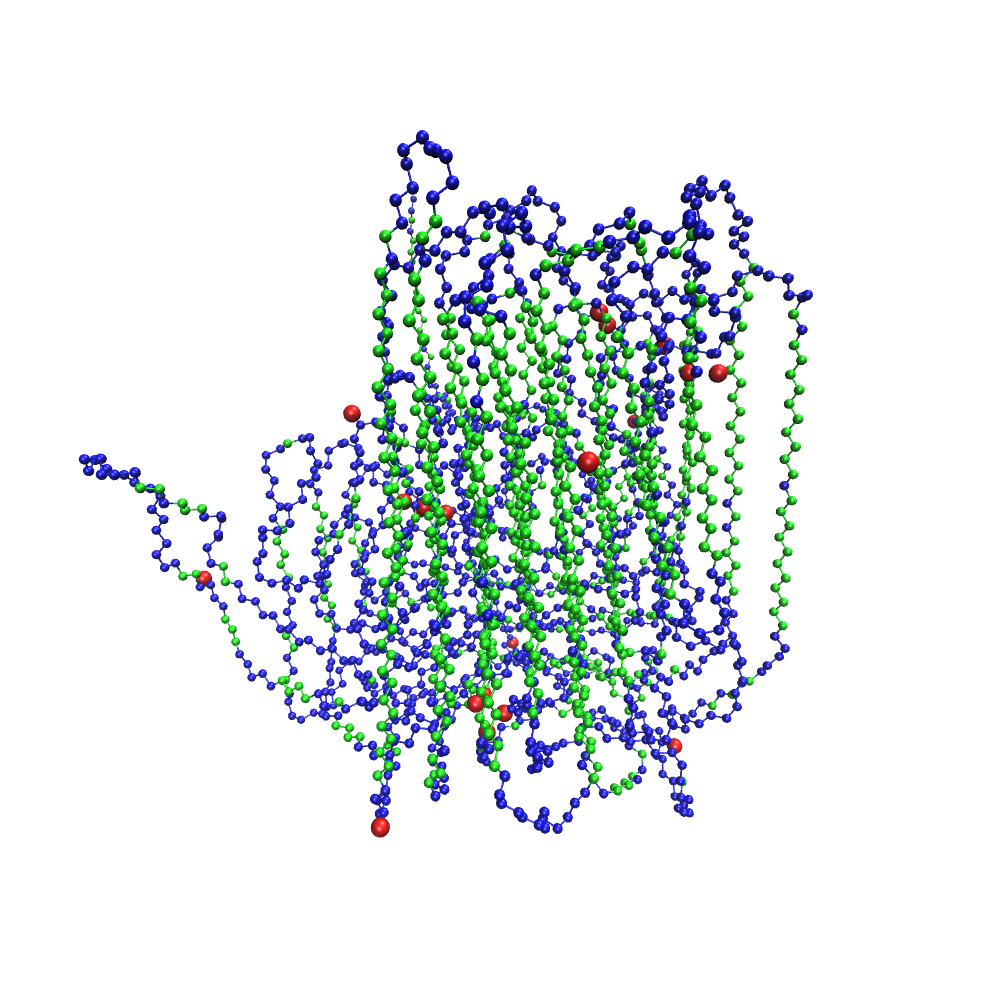}
    \caption{Alcohol}
  \end{subfigure}
  \caption{Equilibrated MD simulation snapshots of (a)~pristine PE and
    PE containing 2~mol\,\% (b)~aldehyde, (c)~ester, (d)~ketone,
    and (e)~alcohol defects.  Carbon backbone atoms are shown in green (crystalline) and blue(amorphous);
    oxygen atoms of the defect groups are highlighted in red.}
  \label{fig:structures}
\end{figure}

Figure~\ref{fig:structures} presents the MD snapshots of equilibrated pristine PE and the four defect-containing systems at 2~mol \,\%. A site order parameter analysis as explained in equation~\eqref{eq:p2_crystallinity} is used to quantify the crystallinity of each structure. Table~\ref{tab:crystallinity} summarizes the crystallinity obtained from order-parameter analysis. In pristine PE, the chains adopt a well-ordered lamellar arrangement  with a global crystallinity of 56.75\% and a density of $0.91 \pm 0.01\text{ g/cm}^3$. At 2~mol\,\%, the ketone defect has the highest crystallinity (50.55\%), followed by ester (39.65\%), aldehyde (36.20\%), and alcohol (31.40\%). In comparison with pristine PE, all defects therefore reduce the crystalline fraction at this concentration. Upon increasing the defect concentration to 4~mol \,\%, crystallinity decreases further for alcohol (19.25\%), ester (26.84\%), and ketone (33.00\%). In contrast, the aldehyde system shows a marked increase to 56.50\%, this behavior is most plausibly explained by the way the aldehyde defect is introduced, where the terminal --CHO group is generated by cleaving the original 2000-unit chain into shorter chains. Thus, the effect of chain-scission causes a decrease in chain length and entanglement that can increase chain mobility and facilitate rearrangement into ordered lamell\ae, so that shorter PE segments can crystallize more readily despite the presence of polar terminal groups \cite{Fayolle_2007,Hsu_2012,Zhang_2020}. The observed crystallinity trends provide a structural context for interpreting the dielectric properties presented in the following sections, since the defect induced changes in morphology directly influence both the magnitude and dynamics of the orientational polarization response.

\begin{table}[htbp]
  \centering
  \caption{Crystallinity of PE systems with different chemical defects
    obtained from MD simulations at 300~K. The semicrystalline PE model without defects has a crystallinity of 56.75\%.}
  \label{tab:crystallinity}
  \begin{tabular}{lcc}
    \toprule
    Defect type & \multicolumn{2}{c}{Crystallinity (\%)} \\
    \cmidrule(lr){2-3}
                & 2~mol\,\% & 4~mol\,\% \\
    \midrule
    Alcohol (--OH)        & 31.40 & 19.25 \\
    Aldehyde (--CHO)      & 36.20 & 56.50 \\
    Ester (--COO--)       & 39.65 & 26.84 \\
    Ketone (\ce{C=O})     & 50.55 & 33.00 \\
    \bottomrule
  \end{tabular}
\end{table}

\subsubsection{Static Dielectric Constant from MD}
\label{sec:md_static}

\begin{figure}[htbp]
  \centering
  \begin{subfigure}[t]{0.48\textwidth}
    \centering
    \includegraphics[width=\linewidth]{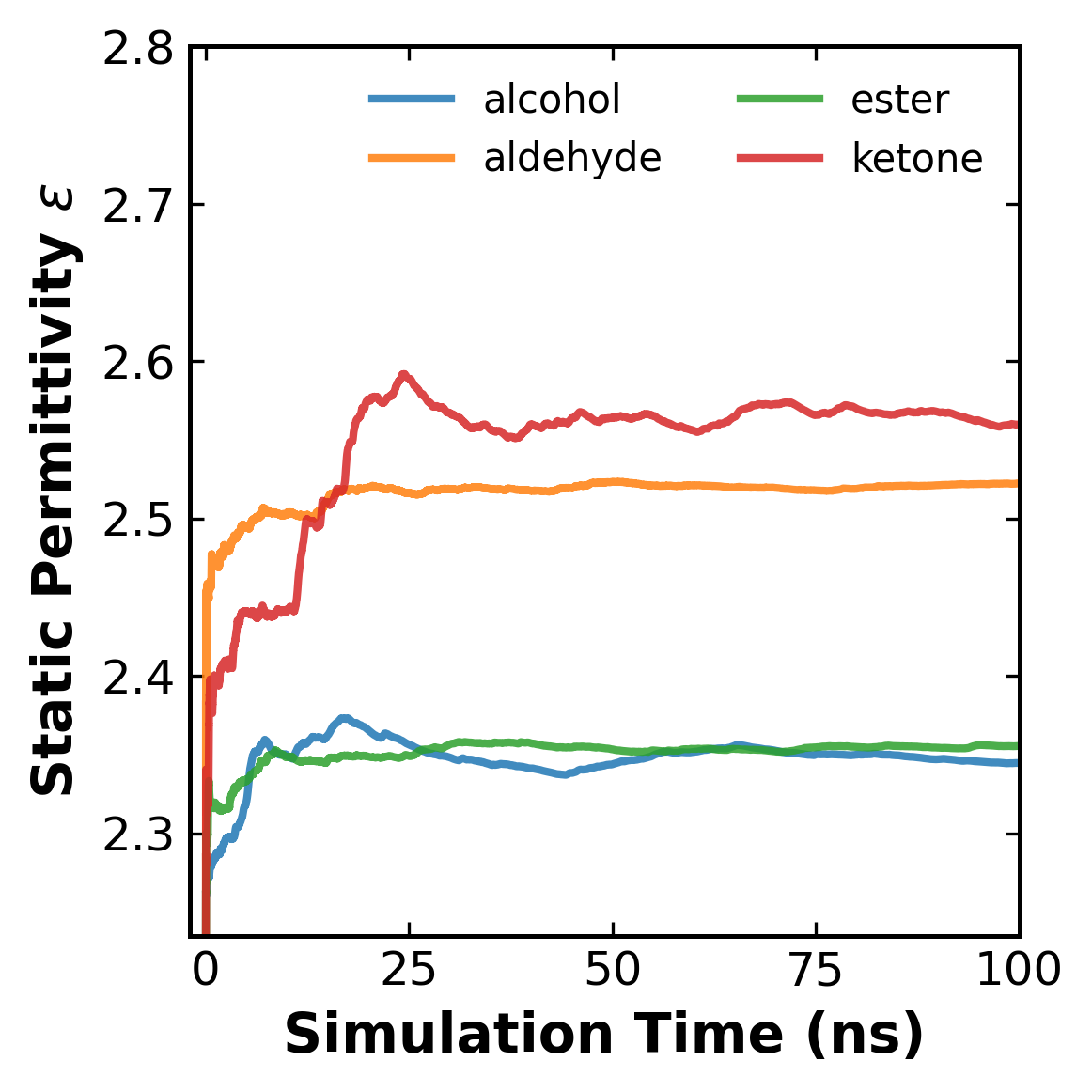}
    \caption{2~mol\,\%}
  \end{subfigure}\hfill
  \begin{subfigure}[t]{0.48\textwidth}
    \centering
    \includegraphics[width=\linewidth]{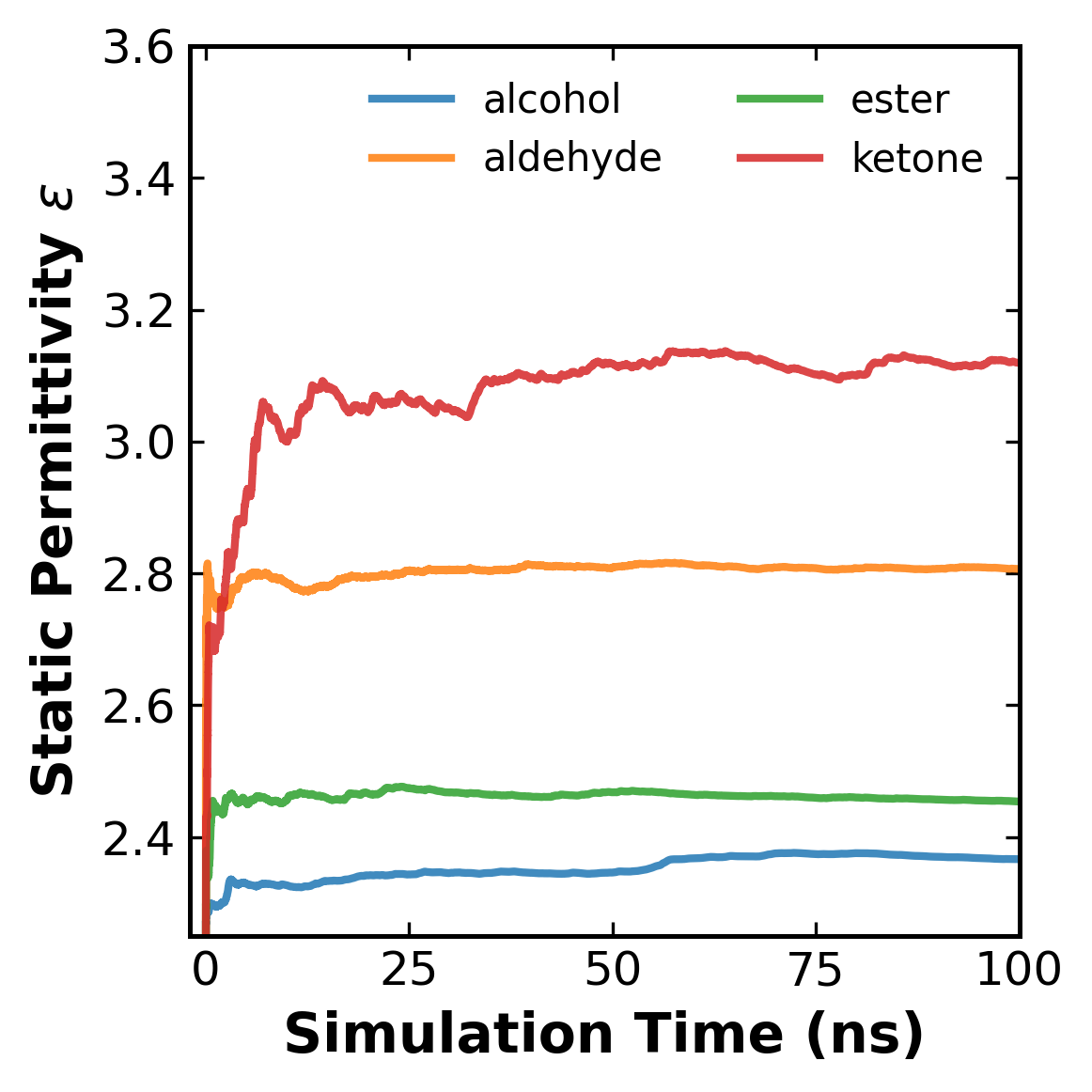}
    \caption{4~mol\,\%}
  \end{subfigure}
  \caption{Convergence of the static dielectric constant $\epss$ as a
    function of simulation time for all four defect types at
    (a)~2~mol\,\% and (b)~4~mol\,\%.}
  \label{fig:eps_conv}
\end{figure}

\begin{table*}[htbp]
  \centering
  \caption{Dielectric properties from MD simulations at 300~K.
    $\epsinf$: high-frequency dielectric constant (from DFPT);
    $\epss$: static dielectric constant (MD);
    $\Deps_\mathrm{tot}$: total dipolar contribution;
    $\Deps_\mathrm{self,def}$, $\Deps_\mathrm{self,mat}$, $\Deps_\mathrm{cross}$:
    self-defect, self-matrix, and cross-correlation contributions}
  \label{tab:md_dielectric}
  \small
  \setlength{\tabcolsep}{4pt}
  \begin{tabular}{ll SSSSS SS}
    \toprule
    {Defect} & {mol\%}
      & {$\epsinf$} & {$\epss$}
      & {$\Deps_\mathrm{self,def}$} & {$\Deps_\mathrm{self,mat}$} & {$\Deps_\mathrm{cross}$} \\
    \midrule
    \multirow{2}{*}{Alcohol}
        & 2 & 2.24 & 2.35 & {+0.104} & {+0.014} & {$-$0.014} \\
        & 4 & 2.26 & 2.40 & {+0.176} & {+0.014} & {$-$0.012} \\
    \addlinespace
    \multirow{2}{*}{Aldehyde}
        & 2 & 2.22 & 2.50 & {+0.245} & {+0.014} & {+0.026}  \\
        & 4 & 2.20 & 2.77 & {+0.505} & {+0.017} & {+0.049}  \\
    \addlinespace
    \multirow{2}{*}{Ester}
        & 2 & 2.22 & 2.33 & {+0.081} & {+0.016} & {+0.019}  \\
        & 4 & 2.20 & 2.43 & {+0.173} & {+0.021} & {+0.033}  \\
    \addlinespace
    \multirow{2}{*}{Ketone}
        & 2 & 2.24 & 2.56 & {+0.247} & {+0.014} & {+0.062}  \\
        & 4 & 2.24 & 3.13 & {+0.686} & {+0.022} & {+0.175}  \\
    \bottomrule
  \end{tabular}
\end{table*}

The static dielectric constant is computed from the equilibrium fluctuations of the total dipole moment M using the Neumann fluctuation formula (equation \eqref{eq:eps0_fluct}) at 300 K. Since non-polarizable force fields neglect electronic polarization by using fixed atomic partial charges, the high-frequency dielectric constant ($\epsinf$), which accounts for the electronic response, was taken from the previous DFPT calculations (Table \ref{tab:dfpt_summary}). This ensures that the dielectric strength ($\Deps = \epss - \epsinf$) obtained from the MD simulations reflects purely the orientational contribution to the static dielectric response. Figure~\ref{fig:eps_conv} shows the time evolution of $\epss$ as a function of simulation time. All systems reach a plateau within 40--80~ns, confirming adequate statistical convergence with minor fluctuations. The converged values are reported in table~\ref{tab:md_dielectric} together with the total dielectric strength ($\Deps$) and the decomposition into defect self-correlations, matrix self-correlations, and cross-correlation contributions (discussed in Section \ref{sec:sc_methd}). 

At 2~mol \,\%, the static permittivity follows the order of, ketone ($\epss = 2.56$) $>$ aldehyde (2.51) $>$ alcohol (2.35) $>$ ester (2.33). At 4~mol \,\% the same order is preserved but the values increase, reaching 3.13 for ketone, aldehyde 2.77, alcohol 2.44, and ester 2.43. The decomposition of the contributions from dipole-fluctuations reveals that the \emph{defect self-correlation} accounts for 70-90\% of the total dielectric response, while the \emph{matrix self-correlation} contributes for 2-14 \%, indicating that the PE backbone segments carry a weak permanent dipole character. At 2 mol \%, the magnitude of $\Deps_{self,def}$ was highest for ketone (+0.247, 76.4\%) and aldehyde (+0.245, 85.7\%), followed by alcohol (+0.105, 99.6\%) and ester (+0.081, 69.7\%), with alcohol displaying a nearly total dominance of the self-defect term. As the concentration increased to 4 mol \%, the ketone maintained the largest absolute value (+0.686, 77.7\%), followed by the aldehyde (+0.505, 88.5\%). Notably, at this higher concentration, ester (+0.173, 76.2\%) surpassed alcohol (+0.138, 97.5\%) in absolute magnitude, though alcohol continued to exhibit the highest relative percentage contribution. The PE-alcohol system exhibits significantly weaker scaling of $\Deps_{\text{self,def}}$ (1.3×) compared to other defect systems. This suggests that, at higher concentrations the molecular mobility of -OH group is constrained, thereby suppressing the orientational contribution to the dielectric response.

The cross-correlation term measures the degree to which the reorientation of one dipole influences the neighboring dipoles, distinguishing between cooperative alignment that increases permittivity and anti-correlated motion that suppresses it. Systems with carbonyl containing defects show positive cross-term contributions, indicating that the reorientation of the C=O dipole induces cooperative alignment of neighboring matrix segments and therefore enhances the total dielectric response beyond the sum of independent self-contributions. Among these, ketone defect system had the strongest cross contributions, with $\Deps_{\text{cross}}$  increasing from  +0.062 at 2 mol \% to +0.175 at 4 mol \%, corresponding to 19.3\% and 19.8\% of the total dielectric strength, respectively. For 
ester defects, the cross-term increases from  $\Deps_{\mathrm{cross}}$ = +0.019 at 2 mol \% to +0.033 at 4 mol \%, corresponding to 16.5\% and 14.6\% of the total response, respectively. Aldehyde defects showed a weaker but still positive cross-term, increasing from +0.026 at  2 mol \% to +0.049 at 4 mol \% and contributing 9.2\% and 8.6\% of the total response, respectively; this reduced coupling is attributed to the terminal –CHO geometry, which allows interaction mainly with only one neighboring chain segment.  In contrast, alcohol defects exhibit negative cross-correlations at both concentrations. At 2 mol \%, $\Deps_{\mathrm{cross}}$ = –0.013, corresponding to –12.4\% of the total response, while at 4 mol \%, $\Deps_{\mathrm{cross}}$ = –0.010 (–7.2\% of the total response). Unlike carbonyl defects, alcohol groups can form hydrogen bonds (O–H···O) with nearby alcohol groups or neighboring PE segments creating a constrained network where the reorientation of a –OH dipole is partially opposed or screened by neighboring dipoles. Thus, this reorientation of alcohol dipoles in a PE chain induces anti-correlated motion in neighboring matrix segments, which suppress the collective dielectric response.
\begin{figure}[htbp]
  \centering
  \includegraphics[width=0.7\textwidth]{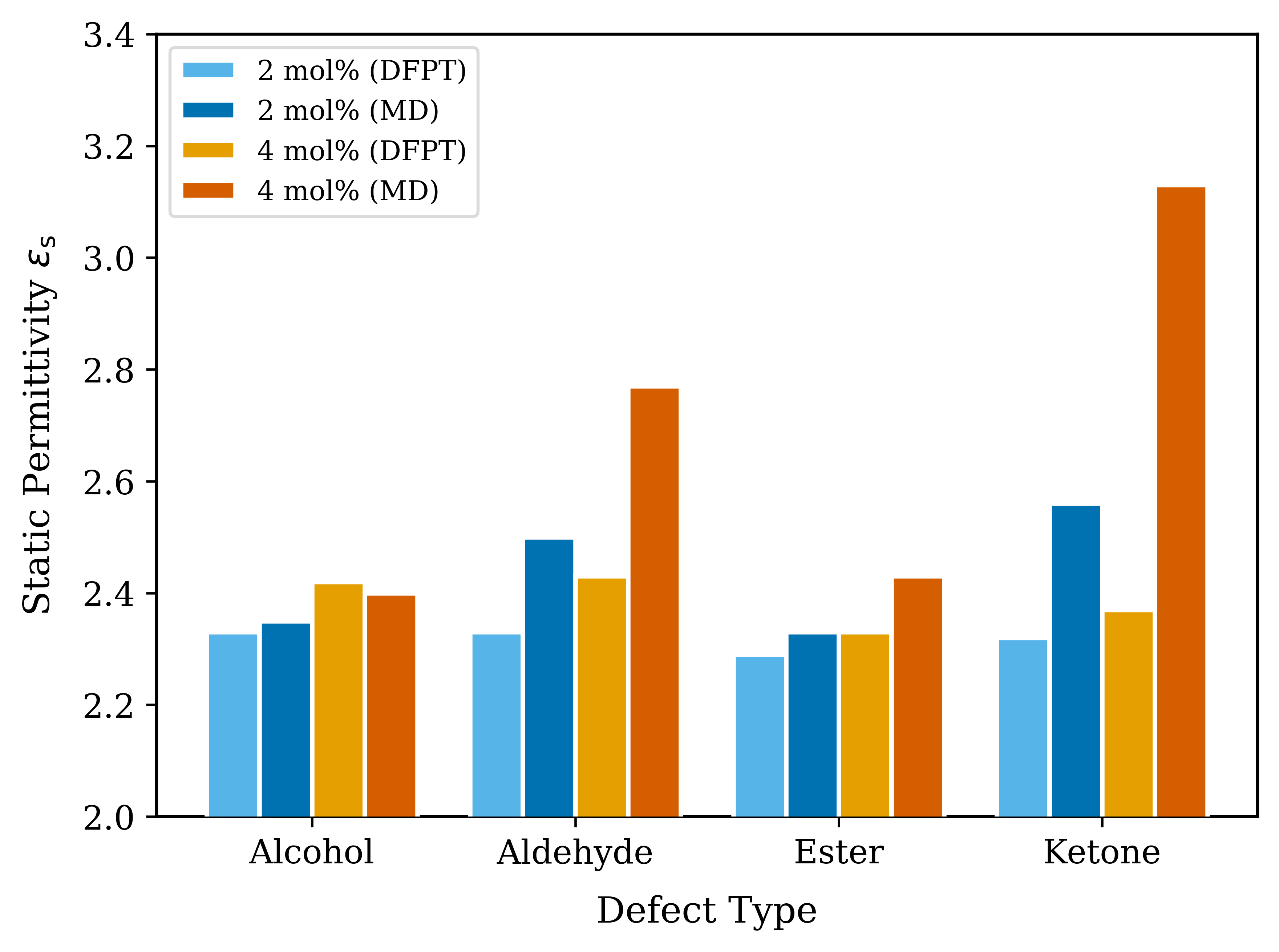}
  \caption{Comparison of the static permittivity from MD simulations and
    DFPT calculations for the four defect types at 2 and 4~mol\,\%.}
  \label{fig:md_dfpt}
\end{figure}

\begin{table}[h!]
\centering
\caption{Static relative permittivity and remnant dipole analysis for defects at 300 K. The permittivity $\epsstilde$ assumes complete dipole relaxation, while $\epss$ excludes the remnant dipole contribution. $\Delta\epsstilde$ is the difference between the two.}
\label{tab:remnant_dipoles}
\begin{tabular}{lcccccc}
\hline
Defect & Conc. & $\epsstilde$ & $\epss$ & $\Delta\epsstilde$ & Remnant  \\
       & (mol\,\%) &  &  & & Fraction (\%) & \\
\hline
Ester  & 2 & 2.37 & 2.35 & 0.02 & 11.5 \\
       & 4 & 2.45 & 2.43 & 0.02 & 9.9  \\[0.1cm]

Alcohol & 2 & 2.36 & 2.35 & 0.01 & 5.5  \\
        & 4 & 2.51 & 2.40 & 0.11 & 35.9  \\[0.1cm]

Ketone  & 2 & 2.61 & 2.56 & 0.05 & 12.7  \\
        & 4 & 3.16 & 3.12 & 0.04 & 3.9  \\[0.1cm]

Aldehyde & 2 & 2.36 & 2.35 & 0.01 & 5.5  \\
         & 4 & 2.77 & 2.77 & 0.00 & 0.00 \\
\hline
\end{tabular}
\end{table}

Figure \ref{fig:md_dfpt} compares the static permittivity $\epss$ calculated from MD dipole fluctuation-dissipation analysis (Eq. \ref{eq:eps0_fluct}) with first-principles DFPT  for all four defect types at 2 and 4 mol \% concentrations. 
For the carbonyl containing defects namely ketone, aldehyde, and ester, the MD-derived static pemittivities are higher than the DFPT values by approximately 5-14\%. This observation is consistent with the additional contribution of orientational polarization arising from the reorientation of C=O dipoles, captured in MD. 
Among these, the ketone system exhibits the largest enhancement, with $\epss$ increasing to 2.56 at 2 mol \% and 3.12 at 4 mol \%, whereas the corresponding DFPT values are 2.32 and 2.37. 
The aldehyde system shows a similar trend ($\epss$ = 2.77 vs. 2.33 at 2 mol \%; 2.77 vs. 2.43 at 4 mol \%), while the ester shows close quantitative agreement within ±2 \% at both concentrations. 
The alcohol system presents a different trend compared to the other defects. The  MD-calculated  $\epss$ values, 2.35 at 2 mol \% and 2.40 at 4 mol \%, are nearly identical to the corresponding DFPT values of 2.33 and 2.42, respectively, despite the presence of a permanent dipole moment.
This behavior can be accounted through the decomposition analysis (Table \ref{tab:md_dielectric}) which reveals that the cross-correlation contribution is negative for alcohol system. This indicates that  the reorientation of alcohol dipoles is effectively hindered by hydrogen-bonding constraints that would otherwise enhance $\epss$ relative to the DFPT prediction.
A second factor is the presence of remnant, or unrelaxed, dipole moments that persist over the MD simulation time scale. This effect is expressed by the ratio $\langle M \rangle^2 / \langle M^2 \rangle$ reported in the last column of Table \ref{tab:remnant_dipoles}.
If we assume that all dipoles are relaxed (for a non ferroelectric polymer) $\langle M \rangle^2=0$ and $\epss=\epsstilde$.
For the carbonyl-containing defects, the remnant fractions remain relatively small, ranging from 3.9 \% for ketone at 4 mol \% to 12.7 \% for ketone at 2 mol \%, indicating that most of the dipole moments relaxes within the simulation window. 
However, the PE system with 4 mol \% alcohol exhibits the largest remnant fraction (35.9 \%). 
This behavior further validate the hydrogen bonding interactions that do not decay over the length of simulation time scale which effectively reduces the static permittivity of the PE-OH systems \cite{misra2014enhanced}.

\subsubsection{Dipolar Relaxation Dynamics}
\label{sec:md_relaxation}

\begin{figure}[htbp]
  \centering
  \begin{subfigure}[t]{0.48\textwidth}
    \centering
    \includegraphics[width=\linewidth]{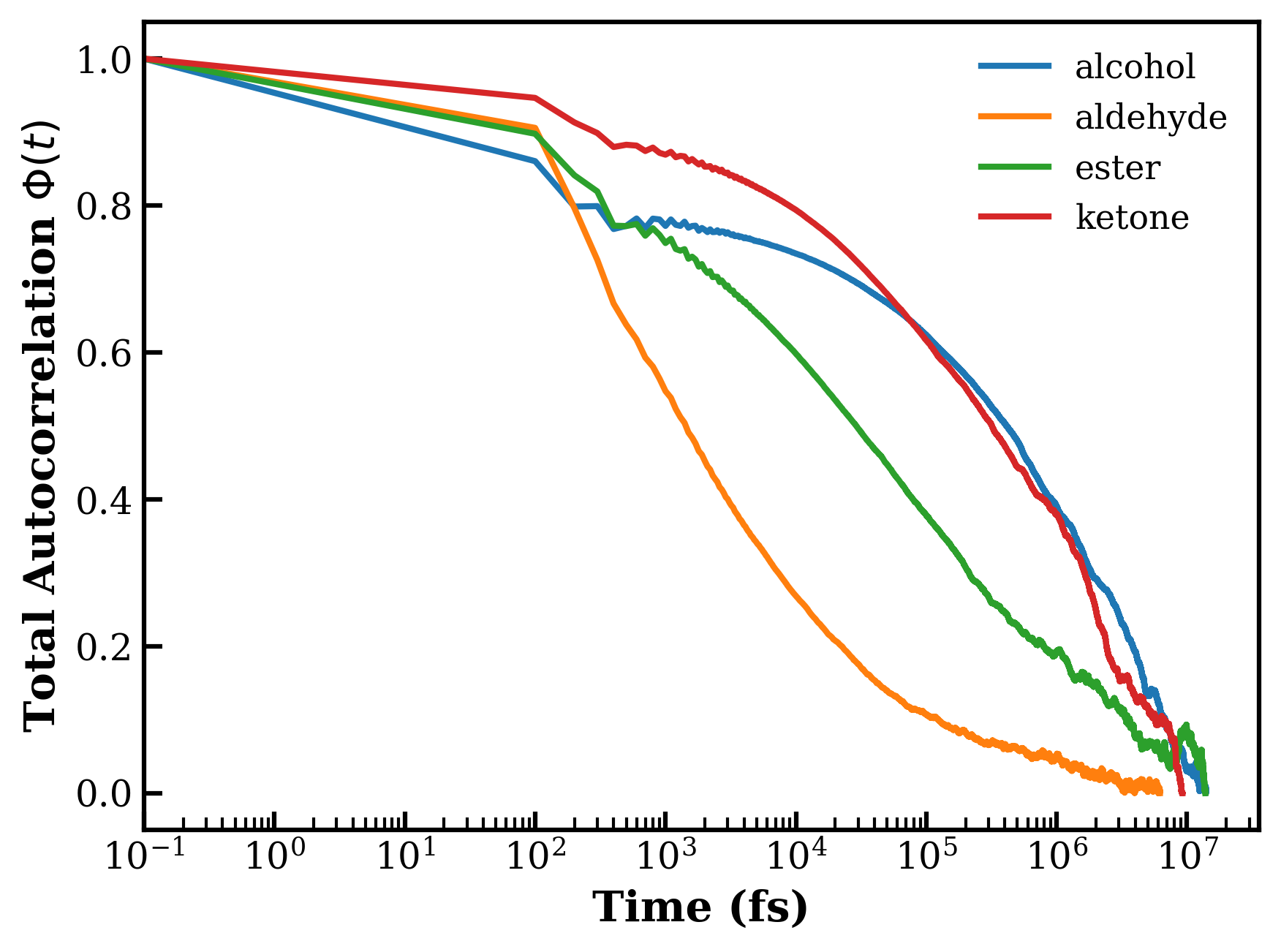}
    \caption{2~mol\,\%}
  \end{subfigure}\hfill
  \begin{subfigure}[t]{0.48\textwidth}
    \centering
    \includegraphics[width=\linewidth]{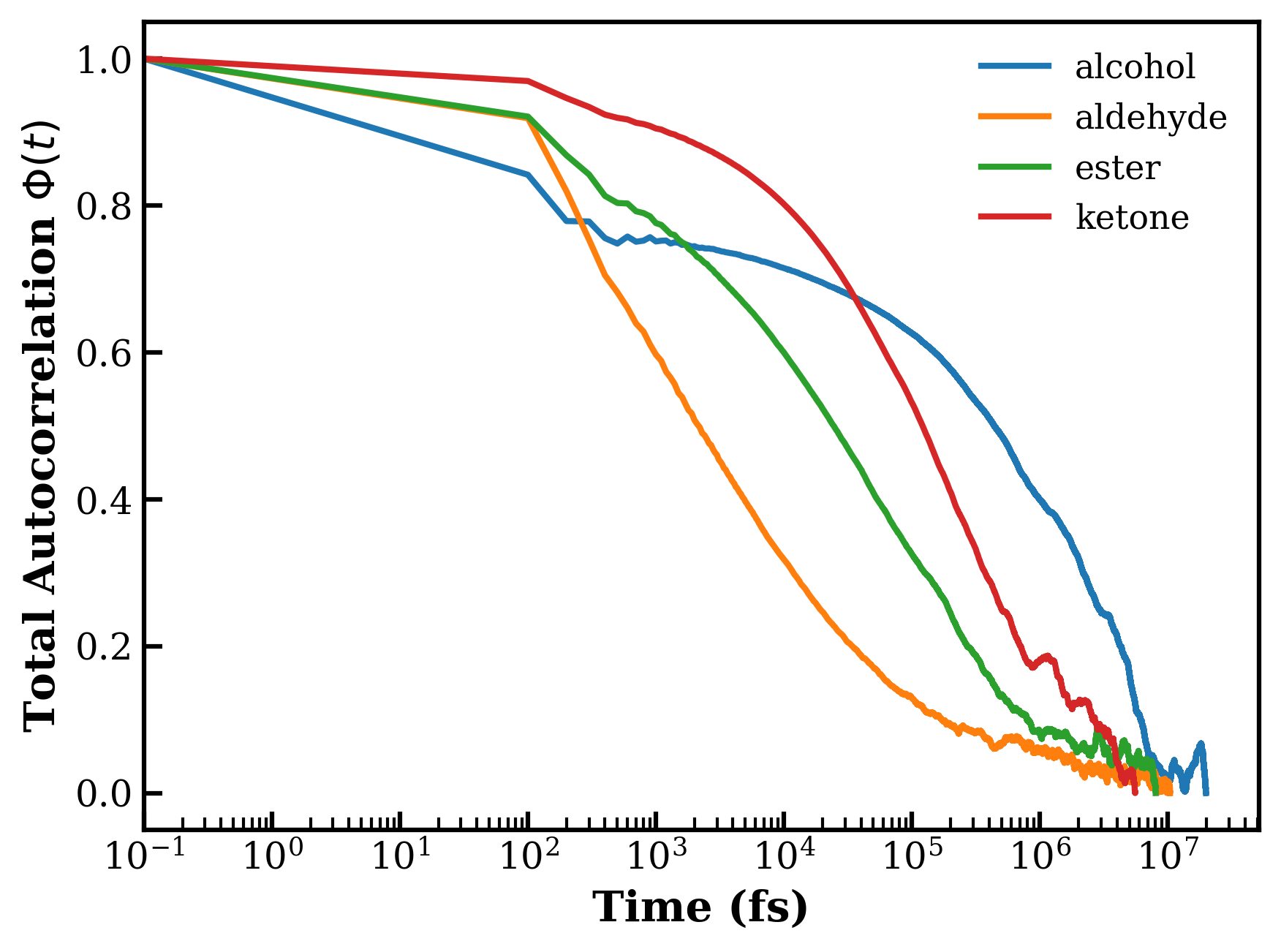}
    \caption{4~mol\,\%}
  \end{subfigure}
  \caption{Normalised total dipole autocorrelation function $\phi(t)$
    for the four defect types at (a)~2~mol\,\% and (b)~4~mol\,\%.}
  \label{fig:acf}
\end{figure}

\begin{figure}[htbp]
  \centering
  \begin{subfigure}[t]{0.48\textwidth}
    \centering
    \includegraphics[width=\linewidth]{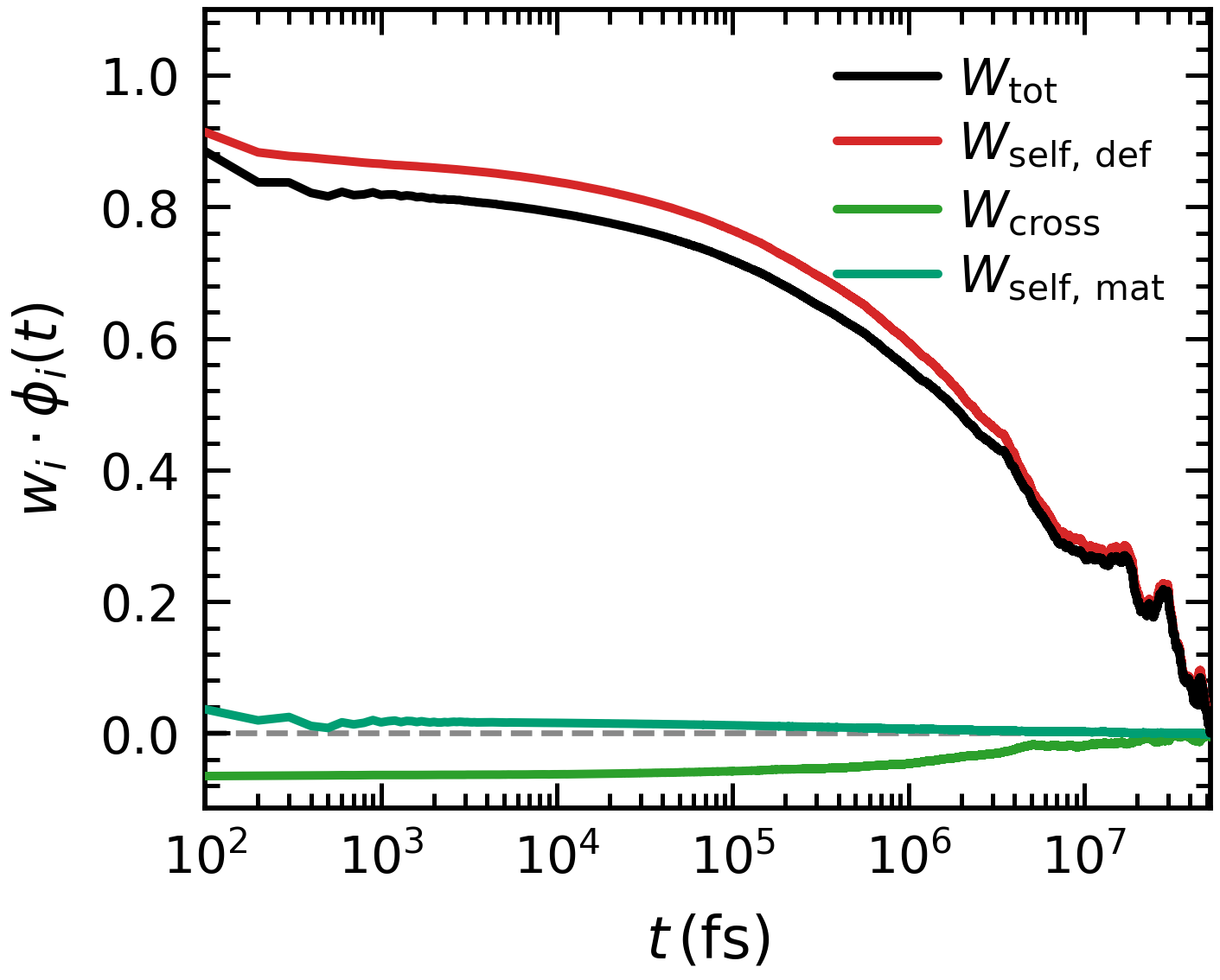}
    \caption{Alcohol}
  \end{subfigure}\hfill
  \begin{subfigure}[t]{0.48\textwidth}
    \centering
    \includegraphics[width=\linewidth]{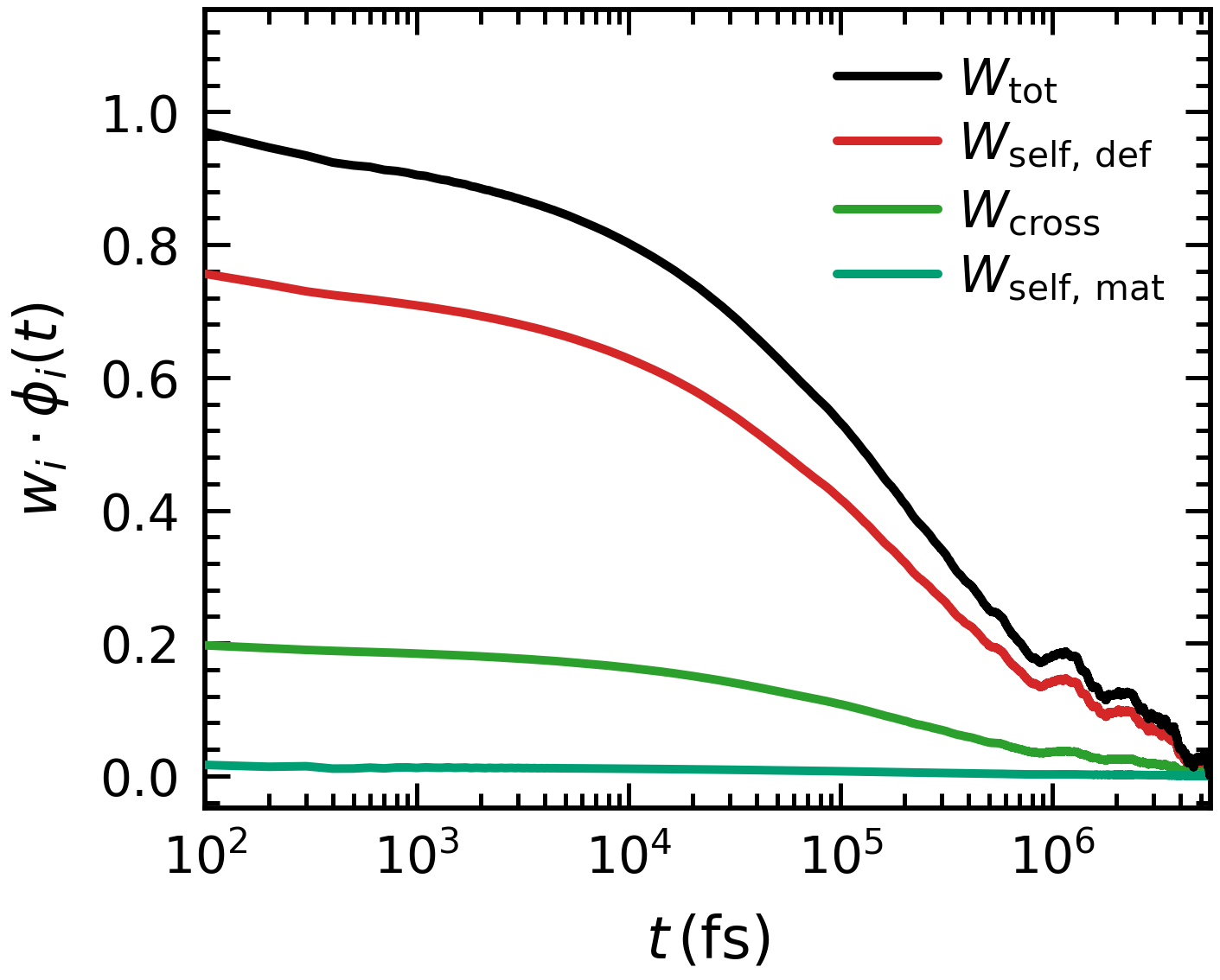}
    \caption{Ketone}
  \end{subfigure}\\[6pt]
  \begin{subfigure}[t]{0.48\textwidth}
    \centering
    \includegraphics[width=\linewidth]{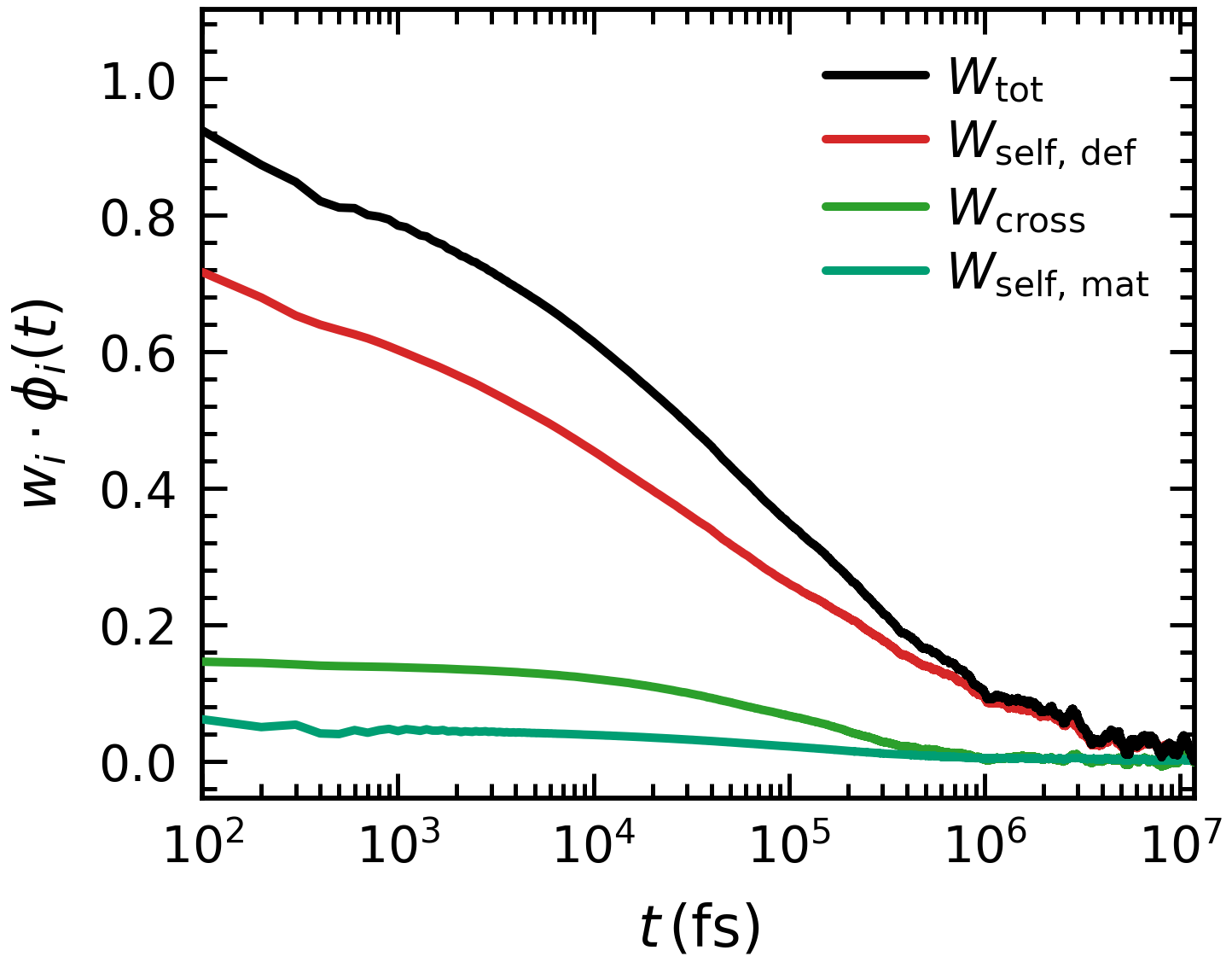}
    \caption{Ester}
  \end{subfigure}\hfill
  \begin{subfigure}[t]{0.48\textwidth}
    \centering
    \includegraphics[width=\linewidth]{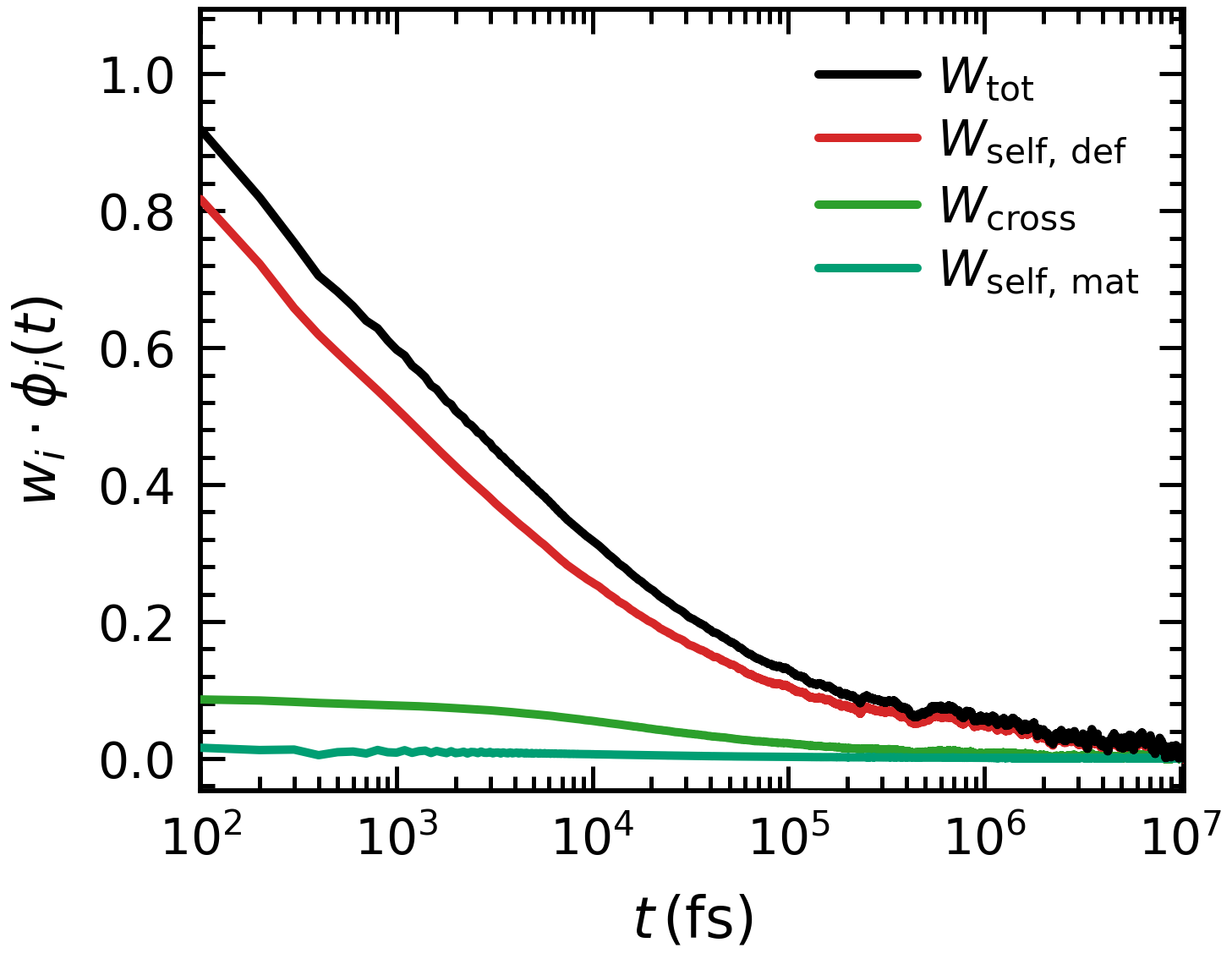}
    \caption{Aldehyde}
  \end{subfigure}
  \caption{Weighted ACF decomposition $w_i \cdot \phi_i(t)$ for the
    4~mol\,\% systems, showing the time-resolved contribution of each
    component to the total normalised ACF.
    (a)~Alcohol, (b)~ketone, (c)~ester, (d)~aldehyde.
    Black: total $\phi_\mathrm{tot}(t)$;
    red: defect self $w_\mathrm{def}\cdot\phi_\mathrm{def}(t)$;
    green: cross $w_\mathrm{cross}\cdot\phi_\mathrm{cross}(t)$;
    cyan: matrix self $w_\mathrm{mat}\cdot\phi_\mathrm{mat}(t)$.
    The weight fractions $w_i = \Deps_i / \Deps_\mathrm{total}$ are derived
    from the static dielectric decomposition (Table~\ref{tab:md_dielectric}).}
  \label{fig:decomp_acf}
\end{figure}

To evaluate the dielectric spectra in the frequency domain, the normalized autocorrelation function of the total dipole moment, $\phi(t)$, was first determined using Equation \ref{eq:acf}. Figure~\ref{fig:acf} shows the normalized dipole ACF $\phi(t)$ for all defect types at both concentrations. $\phi(t)$ comprises contributions from each species and their mutual interactions, thus M(t) was separated into polyethylene-matrix and defect contributions, so the total correlation contains self-terms of each component and a cross-term describing the correlated or anti-correlated dipolar motion between the defect and surrounding the PE-matrix. Figure~\ref{fig:decomp_acf} presents the weighted  ACF components $w_i \cdot \phi_i(t)$ for the 4~mol\,\% systems, where each curve shows  the time-resolved contribution of a given component to the total normalized ACF $\phi_\mathrm{tot}(t)$. This representation directly visualizes how the dielectric  relaxation is built from its constituent parts, with the vertical extent of each curve  reflecting both the static weight ($w_i = \Deps_i / \Deps_\mathrm{total}$) and the time-dependent decay of the corresponding correlation function. The relaxation parameters obtained by fiotting equation \ref{eq:kww}, such as relaxation time $\tau$, and stretching exponent $\alpha$, therefore describe the timescale and distribution of the dipolar relaxation process.

\begin{table}[t]
\centering
\caption{KWW fitting parameters obtained from the normalized dipole autocorrelation functions for defective polyethylene at 300~K. $\tau$ is in nanoseconds.}
\label{tab:decomposition_timescales}
\scriptsize
\setlength{\tabcolsep}{2.5pt}
\renewcommand{\arraystretch}{1.05}

\resizebox{\textwidth}{!}{%
\begin{tabular}{lcccccccccccccccc}
\toprule
Component
& \multicolumn{2}{c}{Es-2}
& \multicolumn{2}{c}{Es-4}
& \multicolumn{2}{c}{OH-2}
& \multicolumn{2}{c}{OH-4}
& \multicolumn{2}{c}{K-2}
& \multicolumn{2}{c}{K-4}
& \multicolumn{2}{c}{Ald-2}
& \multicolumn{2}{c}{Ald-4} \\
\cmidrule(lr){2-3}
\cmidrule(lr){4-5}
\cmidrule(lr){6-7}
\cmidrule(lr){8-9}
\cmidrule(lr){10-11}
\cmidrule(lr){12-13}
\cmidrule(lr){14-15}
\cmidrule(lr){16-17}
& $\tau$ & $\alpha$
& $\tau$ & $\alpha$
& $\tau$ & $\alpha$
& $\tau$ & $\alpha$
& $\tau$ & $\alpha$
& $\tau$ & $\alpha$
& $\tau$ & $\alpha$
& $\tau$ & $\alpha$ \\
\midrule
Total
& 2.49 & 0.26
& 0.77 & 0.29
& 4.52 & 0.55
& 4.47 & 0.56
& 2.52 & 0.58
& 0.91 & 0.40
& 0.34 & 0.19
& 1.10 & 0.17 \\

Def-self
& 2.90 & 0.25
& 0.95 & 0.29
& 4.59 & 0.59
& 4.45 & 0.56
& 2.62 & 0.57
& 0.91 & 0.40
& 0.30 & 0.19
& 1.26 & 0.16 \\

Mat-self
& 3.09 & 0.28
& 0.89 & 0.21
& 6.22 & 0.39
& 2.71 & 0.44
& 1.99 & 0.71
& 0.86 & 0.34
& 0.15 & 0.16
& 1.14 & 0.14 \\

Cross
& 1.75 & 0.33
& 0.21 & 0.47
& 5.07 & 0.82
& 4.00 & 0.57
& 2.25 & 0.65
& 0.85 & 0.36
& 0.50 & 0.22
& 0.55 & 0.21 \\
\bottomrule
\end{tabular}
}

\vspace{0.1cm}
\begin{minipage}{0.95\linewidth}
\footnotesize
Es = ester, OH = alcohol, K = ketone, Ald = aldehyde; 2 and 4 denote defect concentration in mol\,\%.
\end{minipage}
\end{table}

As shown in Table~\ref{tab:decomposition_timescales}, the aldehyde defect has the fastest relaxation ($\tauavg \approx 0.34$~ns at 2~mol\,\%), attributed to the terminal --CHO group that reorients readily at the chain end. It also has a low stretching exponents ($\alpha$ $\approx 0.17$–0.19) indicating a non-exponential relaxation with an extremely broad distribution of timescales. Ketone relaxes more slowly at 2 mol \% (⟨$\tau$⟩=2.52 ns, $\alpha$=0.58) but becomes faster and more heterogeneous at 4 mol \% (⟨$\tau$⟩=0.91 ns, $\alpha$=0.40), suggesting that increased defect concentration accelerates local dipolar relaxation while broadening the relaxation-time distribution. Ester shows a similar shift from 2.49 ns to 0.77 ns with low $\alpha$ values of 0.26–0.29, indicating a distributed non-Debye relaxation.
Among the studied defects, alcohol shows the slowest total dipolar relaxation, with  $\langle \tau \rangle \approx 4.5$~ns at both 2 and 4~mol\,\%. This slow relaxation indicates that the reorientation of hydroxyl dipoles are partially constrained by local O--H$\cdots$O associations.
Reorientation of an alcohol defect therefore requires the rearrangement of the local hydrogen-bonded environment, which delays the decay of the dipole autocorrelation function.\cite{Wang_2013}

\subsubsection{Dielectric Loss Spectra}
\label{sec:loss}

\begin{figure}[htbp]
  \centering
  \begin{subfigure}[t]{0.48\textwidth}
    \centering
    \includegraphics[width=\linewidth]{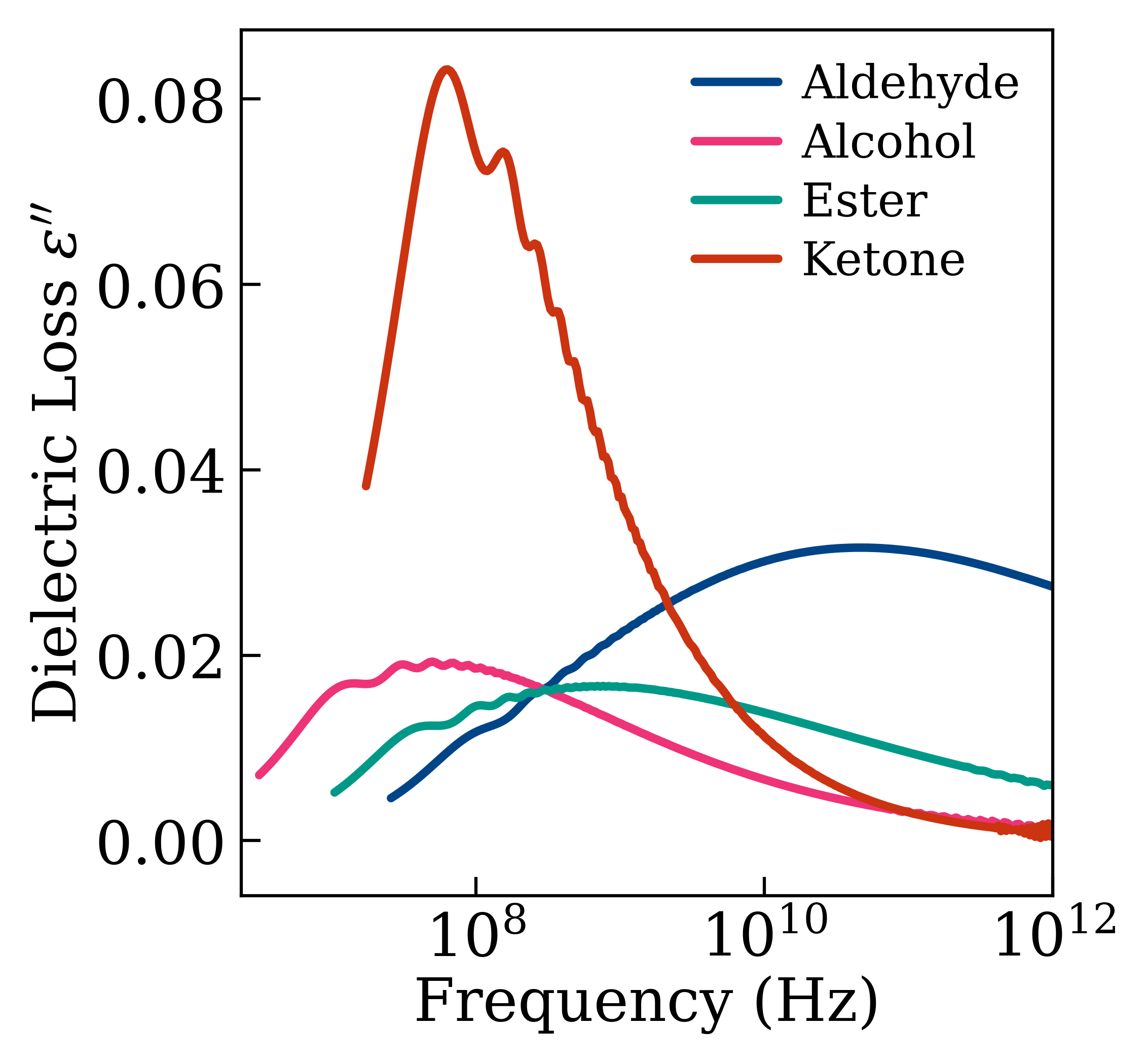}
    \caption{2~mol\,\%}
  \end{subfigure}\hfill
  \begin{subfigure}[t]{0.48\textwidth}
    \centering
    \includegraphics[width=\linewidth]{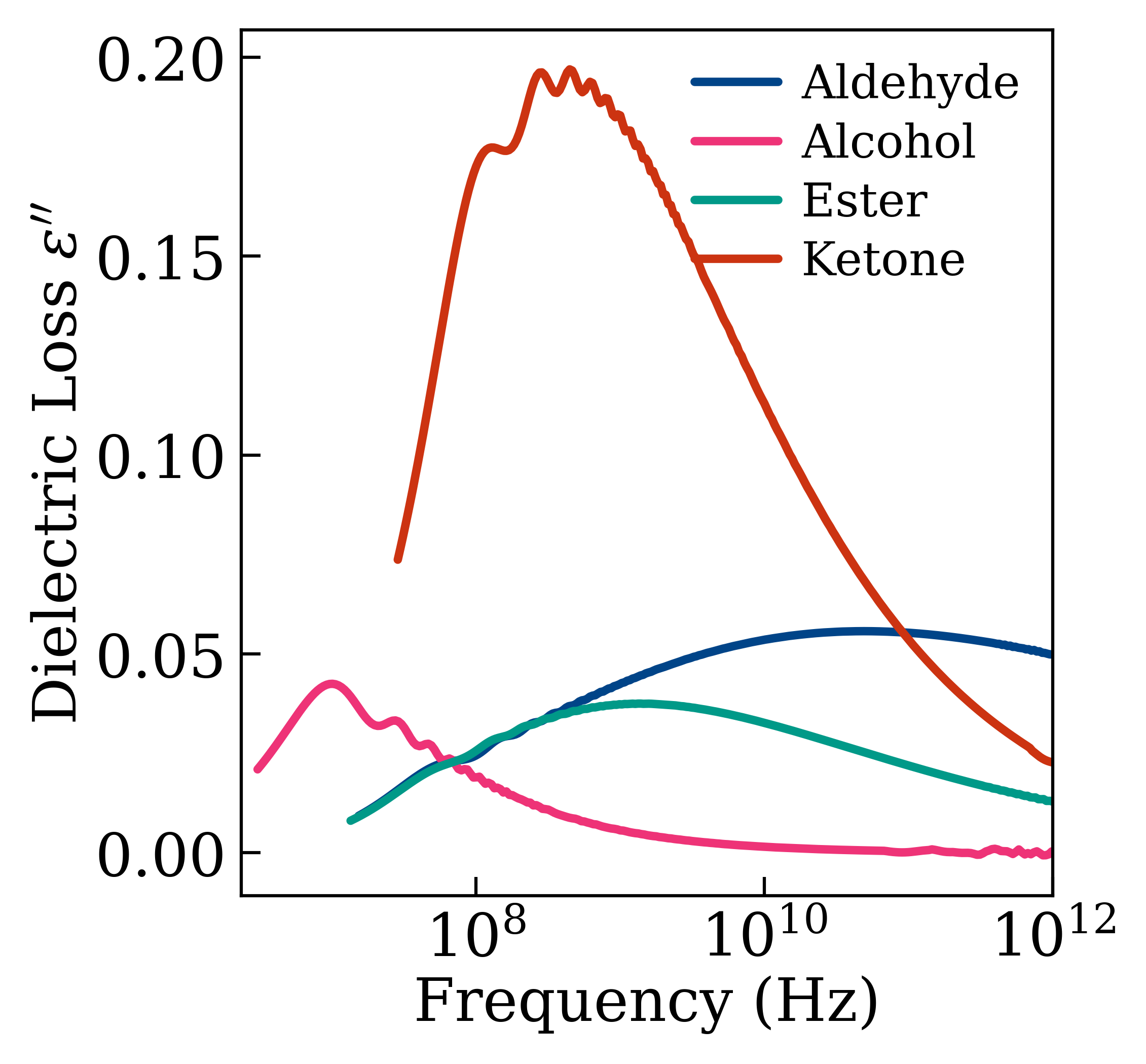}
    \caption{4~mol\,\%}
  \end{subfigure}
  \caption{Total dielectric loss spectra $\epspp(f)$ for all four
    defect types at (a)~2~mol\,\% and (b)~4~mol\,\%.}
  \label{fig:loss}
\end{figure}

The frequency-dependent dielectric loss is obtained by fourier transforming the fitted normalized autocorrelation function, as shown in Figure~\ref{fig:loss}. Across these systems, the loss peak frequencies exhibit a broad distribution, reflecting fundamentally different relaxation mechanisms across different defect types. Aldehyde relaxes at the highest frequencies, with maxima at 45.3 GHz and 50.1 GHz for 2 and 4~mol\,\%, respectively, indicating the fastest dipolar relaxation among the four defects. Ester shows intermediate-frequency loss peaks at 689 MHz for 2~mol\,\% and 1.49 GHz for 4~mol\,\%, while alcohol remains at the lowest-frequency peak position, 31 MHz, for both concentrations, consistent with slow hydroxyl-mediated relaxation. The maximum dielectric loss increases with defect concentration for all systems. Among the 2 mol\% systems, ketone shows the highest loss amplitude, with $\varepsilon''_{\max}$ = 0.083, followed by aldehyde (0.032), alcohol (0.024), and ester (0.017). At 4 mol\%, ketone again gives the largest loss, increasing to $\varepsilon''_{\max}$= 0.19, followed by aldehyde (0.056), ester (0.038), and alcohol (0.034).

\begin{figure}[htbp]
  \centering
  \begin{subfigure}[t]{0.24\textwidth}
    \centering
    \includegraphics[width=\linewidth]{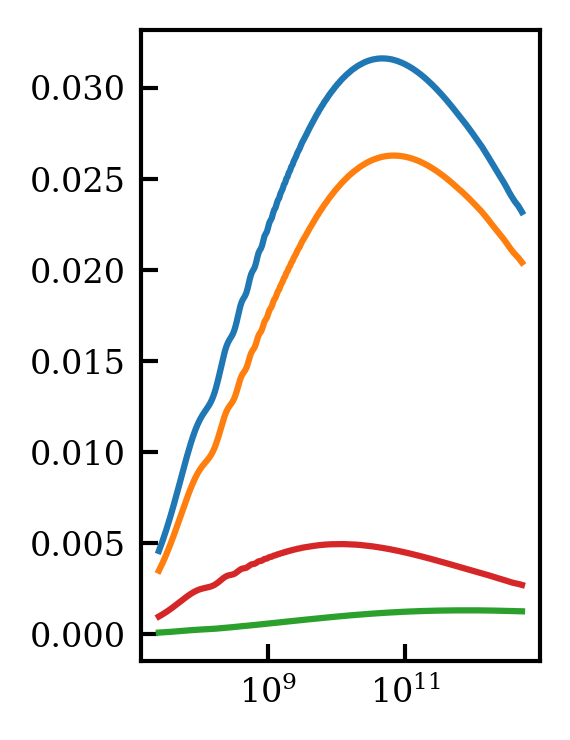}
    \caption{Ald.\ 2\%}
  \end{subfigure}\hfill
  \begin{subfigure}[t]{0.24\textwidth}
    \centering
    \includegraphics[width=\linewidth]{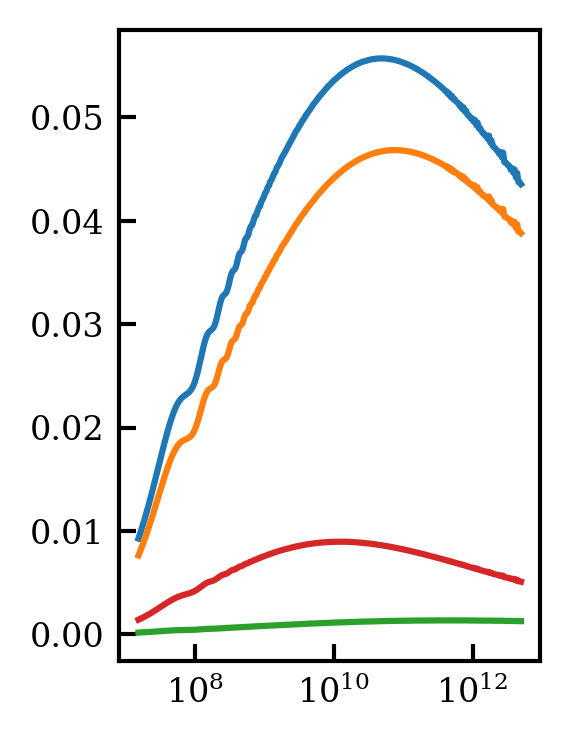}
    \caption{Ald.\ 4\%}
  \end{subfigure}\hfill
  \begin{subfigure}[t]{0.24\textwidth}
    \centering
    \includegraphics[width=\linewidth]{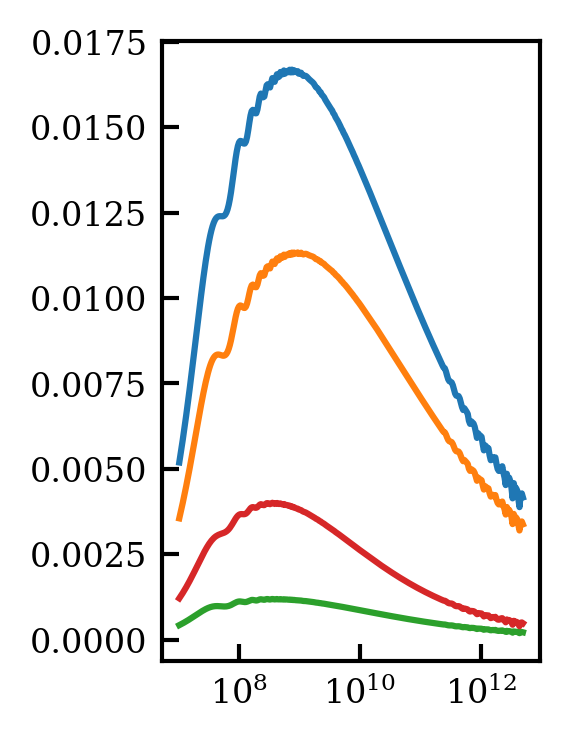}
    \caption{Est.\ 2\%}
  \end{subfigure}\hfill
  \begin{subfigure}[t]{0.24\textwidth}
    \centering
    \includegraphics[width=\linewidth]{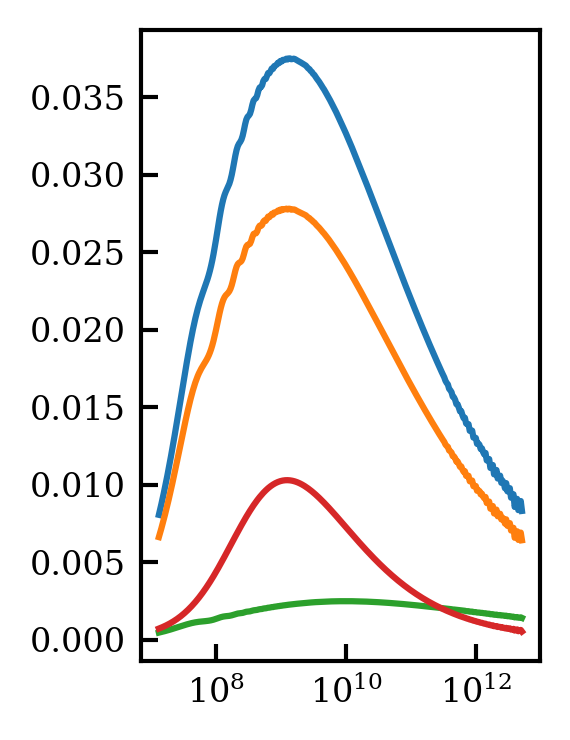}
    \caption{Est.\ 4\%}
  \end{subfigure}\\[6pt]
  \begin{subfigure}[t]{0.24\textwidth}
    \centering
    \includegraphics[width=\linewidth]{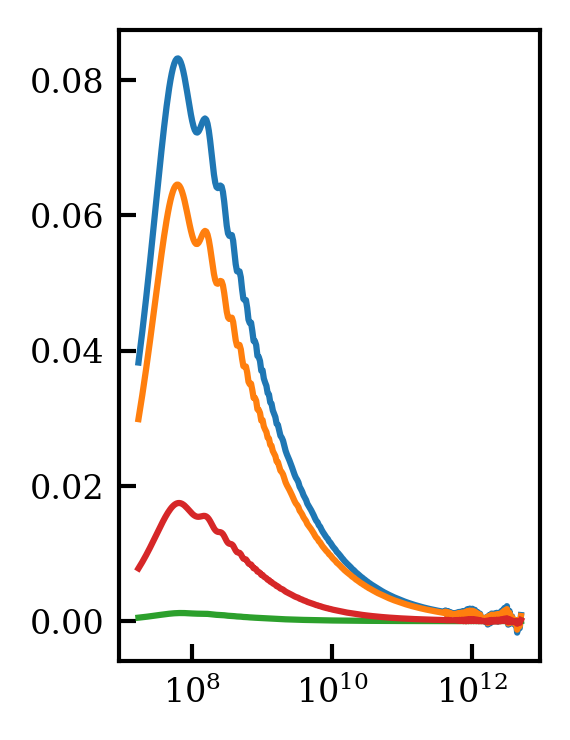}
    \caption{Ket.\ 2\%}
  \end{subfigure}\hfill
  \begin{subfigure}[t]{0.24\textwidth}
    \centering
    \includegraphics[width=\linewidth]{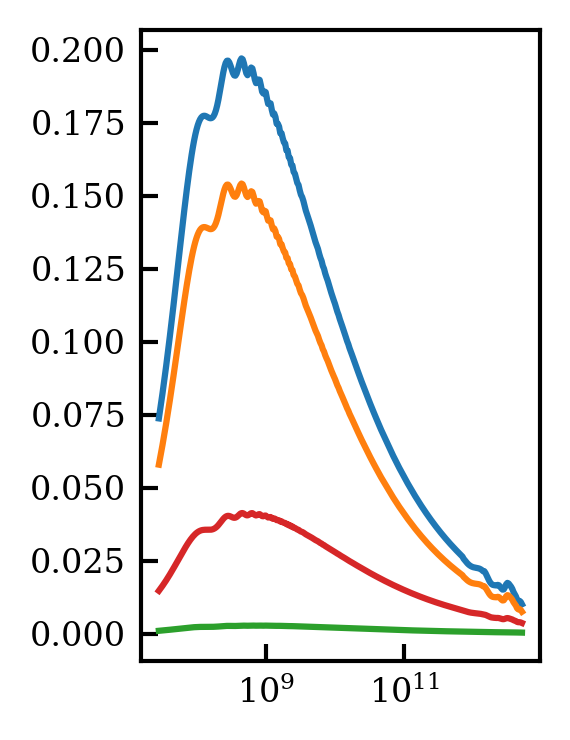}
    \caption{Ket.\ 4\%}
  \end{subfigure}\hfill
  \begin{subfigure}[t]{0.24\textwidth}
    \centering
    \includegraphics[width=\linewidth]{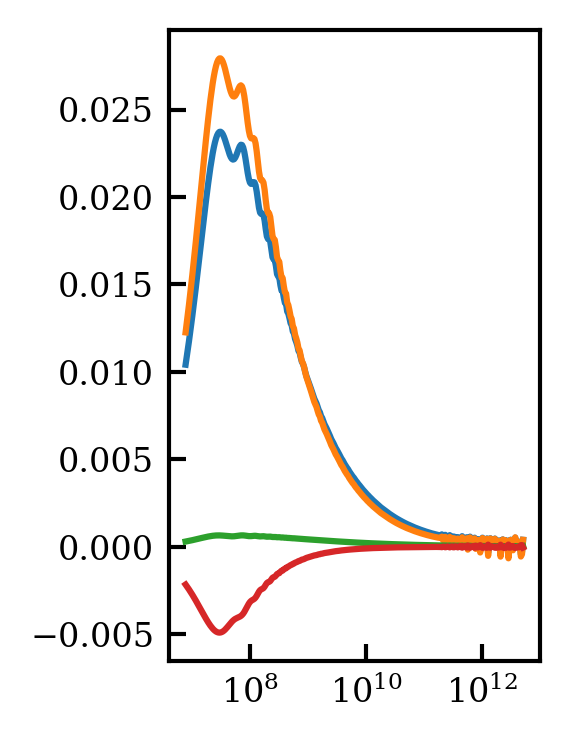}
    \caption{Alc.\ 2\%}
  \end{subfigure}\hfill
  \begin{subfigure}[t]{0.24\textwidth}
    \centering
    \includegraphics[width=\linewidth]{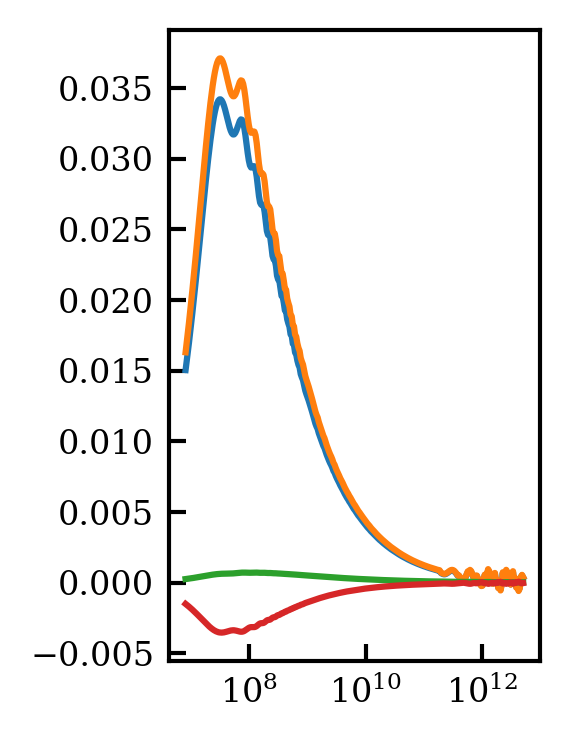}
    \caption{Alc.\ 4\%}
  \end{subfigure}
  \caption{Decomposition of the dielectric loss spectrum into self (defect),
    self (matrix), and cross contributions for (a,b)~aldehyde, (c,d)~ester,
    (e,f)~ketone, and (g,h)~alcohol at 2 and 4~mol\,\%.
    Blue: total; orange: defect self; red: matrix self; green: cross sum.}
  \label{fig:loss_decomp}
\end{figure}

To better understand the origin of the each system’s dielectric response, we decompose the total loss into contributions of the self and cross interactions, as shown in Figure~\ref{fig:loss_decomp}. For aldehyde, ester, and ketone systems, the defect self-term dominates the loss spectrum with the total closely following the defect contribution, while positive cross-correlations provide moderate enhancement visible as green curves that peak at frequencies slightly shifted from the main loss maximum.  For the ketone defect, the cross-correlation contribution is large, especially at 4~mol\,\%, where $\Deps_\mathrm{cross} = +0.175$ constitutes nearly 20\% of the total $\Deps$. In contrast, alcohol systems (panels g, h) exhibit negative cross-correlation contributions (green curves dipping below zero in the 10–100 MHz range), directly demonstrating the anti-correlated suppression mechanism that reduces the total dielectric loss\cite{borodin2009}. The decomposed spectra reveal that cooperative enhancement in carbonyl systems arises primarily at frequencies near the main loss peak, with ketone showing the strongest cross-correlation enhancement consistent with its +19.3\% to +19.8\% fractional contribution to the static response. Matrix self-contributions remain negligible across all systems, confirming that dielectric loss originates almost entirely from defect incorporation rather than intrinsic polyethylene polarization.  The  spectra reveal that cooperative enhancement in carbonyl systems arises primarily at frequencies near the main loss peak ($10^9$--$10^{10}$~Hz), while alcohol's negative cross-correlation arises most strongly at lower frequencies ($\approx 10^8$--$10^9$~Hz) where hydrogen-bond network rearrangements dominate. Overall, the results establish a direct molecular interpretation of the dielectric spectra of polyethylene with defects and highlight the central role of different polarization mechanisms in shaping its macroscopic dielectric behavior.

\section{Conclusions}
\label{sec:conclusions}

We have presented a multiscale computational study of the dielectric response of semi-crystalline polyethylene containing seven types of chemical defects---alcohol, aldehyde, ketone, carboxylic acid, enone, ester, and vinylene---at concentrations of 2 and 4~mol\,\%, combining first-principles DFPT calculations with classical molecular dynamics simulations.

The DFPT calculations on a lamellar slab model reveal that all polar defects increase the static dielectric constant in the low temperature limit, $\epszero$, relative to pristine PE, even when some of them (aldehyde, ester, carboxyl) suppress the high-frequency electronic component $\epsinf$. Mode-resolved analysis of the phonon spectrum shows that this enhancement is driven by the $1/\omega_m^2$-weighted contributions of defect-activated IR-active modes. Aldehyde yields the largest ionic increment ($\Delta\eps = 0.228$ at 4~mol\,\%) because 69\% of its dielectric mode weight resides below 100~cm$^{-1}$, despite carrying the smallest carbonyl-related Born effective charges. Ester, which has the largest Born charges, achieves only a moderate increment because its polar modes are stiffer. Vinylene, the only non-polar defect, produces a negligible shift in $\epszero$, with its mode weight concentrated at high frequencies. These results establish that the static dielectric response of defected PE is governed by the interplay between dynamical charges and phonon softness, not by either factor alone.
The MD simulations extend this picture to finite temperature and capture the orientational polarization dynamics inaccessible to DFPT. The ketone defect produces the largest total static permittivity ($\epss = 3.13$ at 4~mol\,\%) and dielectric loss, driven by a strong positive cross-correlation between defect and matrix dipoles ($\Deps_\mathrm{cross} = +0.175$). The alcohol defect is the slowest to relax ($\tauavg \approx 4.46$~ns at 4~mol\,\%) due to hydrogen bonding interactions, and is the only defect exhibiting a negative cross-correlation that partially screens its dielectric contribution. The aldehyde defect relaxes fastest ($\tauavg \approx 0.34$~ns) owing to the compact terminal --CHO group. Decomposition of the dielectric loss spectra into defect-self, matrix-self, and cross terms provides a quantitative assignment of each component's spectral contribution, showing that the defect self-correlation dominates in all cases but that inter-component coupling can account for up to 20\% of the total response.

\begin{acknowledgments}
  Simone Vincenzo Suraci, Davide Fabiani and Josy Cohen are gratefully acknowledged for interesting discussions in the early stages of this work.
  This work was granted access to the HPC resources of TGCC and IDRIS under the allocations 2026-A0190906018, 2025-A01709
06018, 2024-A0150906018 made by GENCI.
\end{acknowledgments}

\bibliography{references}

\end{document}